\documentclass[twocolumn,showpacs,preprintnumbers,a4,prb,superscriptaddress]{revtex4}
\usepackage{graphicx}%
\usepackage{dcolumn}
\usepackage{amsmath}
\usepackage{amssymb}
\usepackage{bm}

\begin{document}
\title{Crystal and magnetic structure of La$_{1-x}$Ca$_{x}$MnO$_{3}$ compound $(0.11\leq x\leq 0.175)$}
\author{M. Pissas\cite{cor}}
\affiliation{Institute of Materials Science, National Center for
Scientific Research, Demokritos, 15310 Ag. Paraskevi, Athens,
Greece}
\author{I. Margiolaki}
\affiliation{European Synchrotron Radiation Facility (ESRF), Pluo
30.1.08 6 Rue J. Horowitz, 38043 Grenoble, Cedex 9, France}.
\author{G. Papavassiliou}
\author{D. Stamopoulos}
\affiliation{Institute of Materials Science, National Center for
Scientific Research, Demokritos, 15310 Ag. Paraskevi, Athens,
Greece}
\author{D. Argyriou}
\affiliation{Berlin Neutron Scattering Center,
Hahn-Meitner-Institut, Glienicker Str. 100 D-14109, Berlin,
Germany}

\date{\today }

\begin{abstract}
We studied the crystal and magnetic structure of the
La$_{1-x}$Ca$_{x}$MnO$_{3}$ compound for $(0.11\leq x\leq 0.175)$
using stoichiometric samples. For $x<0.13$ the system's ground
state is insulating canted antiferromagnetic. For $0.13\le x \le
0.175$ e below the Jahn Teller transition temperature ($T_{\rm
JT}$) the crystal structure undergoes a monoclinic distortion. The
crystal structure can be described with $P 2_1/c$ space group
which permits two Mn sites. The unit cell strain parameter
$s=2(a-c)/(a+c)$ increases for $T<T_{\rm JT}$, taking the maximum
value at the Curie point, and then decreases. Below $T_{\rm
M^/M^{//}}\approx 60 $ K $s$ abruptly changes slope and finally
approaches $T=0$ K with nearly zero slope. The change of $s$ at
$T_{\rm M^/M^{//}}$ is connected to a characteristic feature in
the magnetic measurements. As $x$ increases towards the
ferromagnetic metallic boundary, although $s$ is reduced
appreciably, the monoclinic structure is preserved. The monoclinic
structure is discussed with relation to the orbital ordering,
which can produce the ferromagnetic insulating ground state. We
also studied samples that were prepared in air atmosphere. This
category of samples shows ferromagnetic insulating behavior
without following the particular variation of the $s$ parameter.
The crystal structure of these samples is related to the so-called
O$^{*}$ ($c>a>b/\sqrt{2}$) structure.
\end{abstract}

\pacs{75.47.Lx,75.47.GK,75.25.+z,75.30Cr} \maketitle

\section{Introduction} The nature of the ferromagnetic insulating
state (FMI) in lightly doped La$_{1-x}$A$_x$MnO$_3$ (A=Sr, Ca,
$x=0.1-0.2$) rare-earth manganite perovskites continues to offer
significant challenges in our understanding of the physical
properties of these materials. Typically, insulating compounds of
this category are antiferromagnetic, and ferromagnetism is
associated with a metallic behavior that arises from doping and
double exchange. The FMI state can be realized with a certain
orbital ordering favorable for ferromagnetism.\cite{khomskii97} In
the intermediate doping levels ($x=0.23-0.4$) where the
ferromagnetic metallic (FMM) state is observed the double-exchange
mechanism is able to explain some of the experimental results.
Despite of the numerous studies concerning the FMI phase of
La$_{1-x}$Sr$_x$MnO$_3$ compound, for the Ca case only a few
exist. Furthermore, the tendency of this system in particular to
produce cation deficient samples complicates the situation. For
the Sr case the $x=0.125$ sample shows a transition to a FMM state
from paramagnetic insulating state, followed by an abrupt
first-order transition to the FMI phase. The FMI state coexists
with a weak antiferromagnetic component with the $A-$type magnetic
structure as was found in LaMnO$_3$.\cite{pinsard97} In addition
in some other studies\cite{yamada00,endoh99} charge or orbital
ordering have been proposed to occur in the FMI state.

\begin{figure}[htbp] \centering%
\includegraphics[angle=0,width=1.0 \columnwidth]{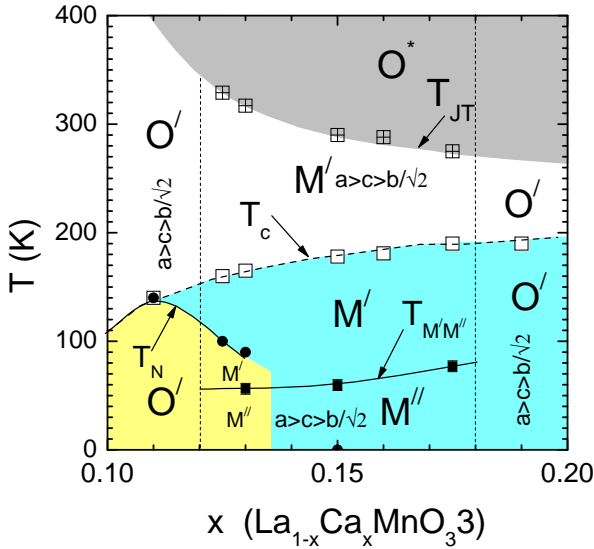}
\caption{Ferromagnetic insulating part of the phase diagram of
La$_{1-x}$Ca$_{x}$MnO$_{3}$ compound. The light yellow region
corresponds to the CAF regime, and the light blue to the FMI
regime.  The symbols O*, O$^/$, O$^{//}$, M$^/$ and M$^{//}$ have
the following meaning: [O* orthorhombic $Pnma$ symmetry with
$c\geq a\geq b/\sqrt{2}$],[O$^/$,(O$^{//}$) orthorhombic $Pnma$
symmetry with $a>c>b/\sqrt{2}$ and strong (intermediate)
anisotropic bond lengths] and [M$^/$, (M$^{//}$) monoclinic $P
2_1/c$ symmetry with $a>c>b/\sqrt{2}$ and strong (intermediate)
anisotropic bond lengths]. The $T_{\rm M^/M^{//}}$ is defined at
the temperature where the slope of the $a(T)$ and $b(T)$ curves
change slope abruptly upon heating from $T=0$. The $T_{\rm JT}$
curve is defined at the temperature where the $m(T)$ curves
display a jump-like or slope-change behavior. The $T_{\rm C}$ and
$T_{\rm N}$ are both estimated  from neutron and magnetization
data.}
\label{diagram}%
\end{figure}%

\begin{figure}[htbp] \centering%
\includegraphics[angle=0,width=0.7 \columnwidth]{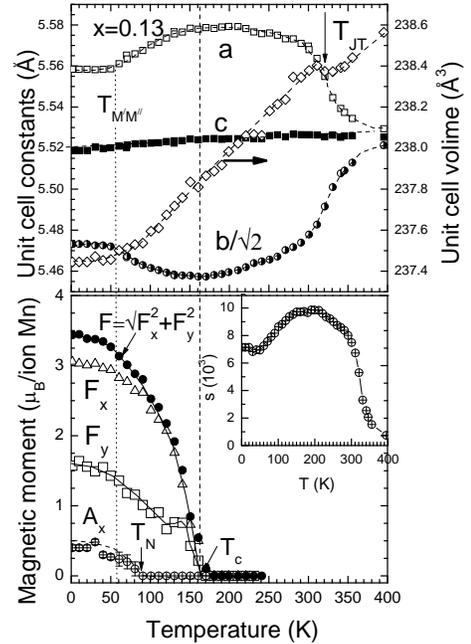}
\caption{ (upper panel) Temperature variation of the lattice
parameters and unit-cell volume for
La$_{0.87}$Ca$_{0.13}$MnO$_{3}$ compound. (lower panel)
Temperature variation of the ordered ferromagnetic and
antiferromagnetic moments. The ferromagnetic components are
denoted by $F$ and the antiferromagnetic with $A$. The inset shows
the temperature variation of the spontaneous orthorhombic strain
$(s=2(a-c)/a+c))$. Lines through the experimental points are
guides to the eye.}
\label{abc-r013}%
\end{figure}%

\begin{figure}[tbp] \centering%
\includegraphics[angle=0,width=0.7 \columnwidth]{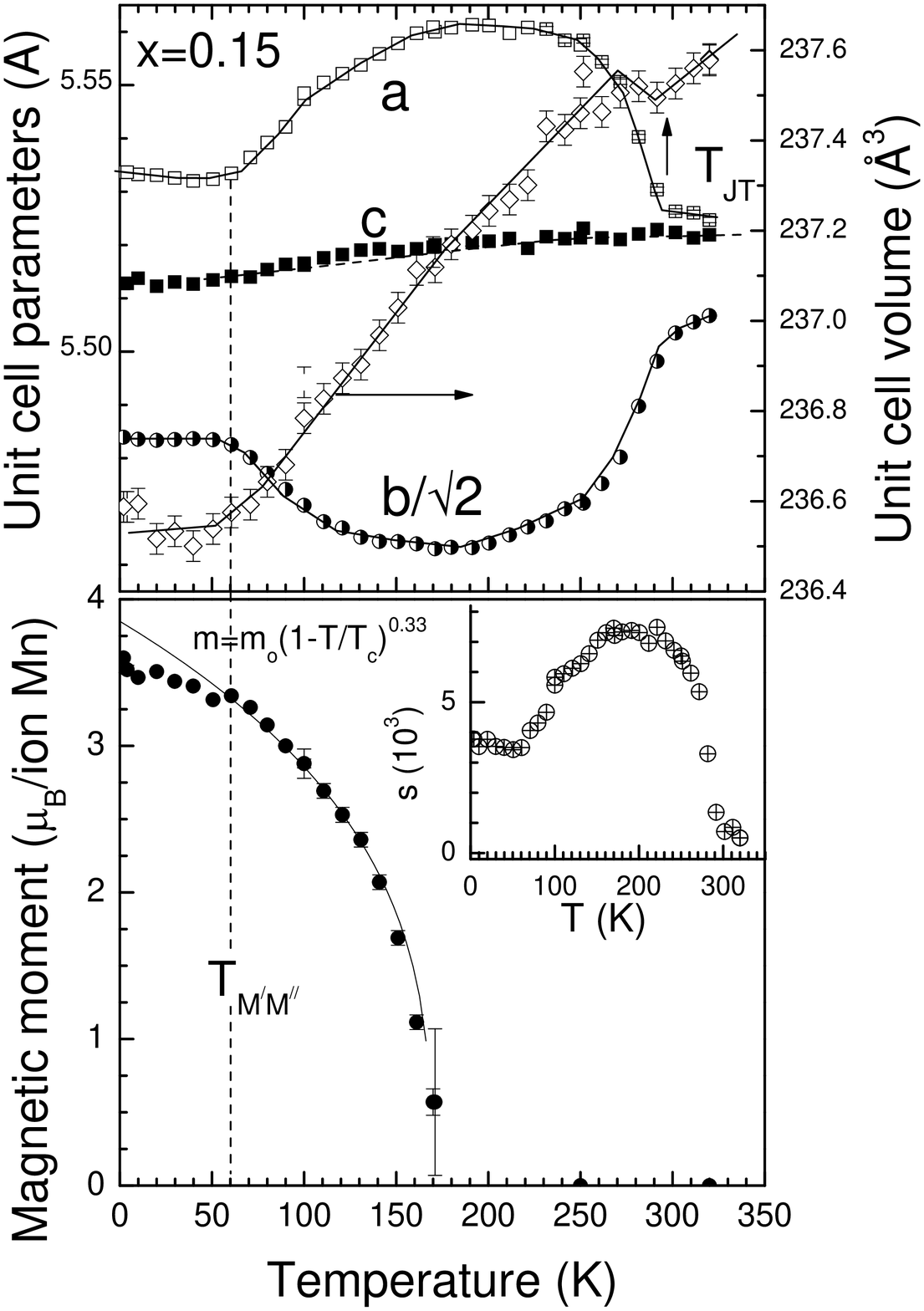}
\caption{ (upper panel) Temperature variation of the lattice
parameters and unit-cell volume for
La$_{0.85}$Ca$_{0.15}$MnO$_{3}$ compound. Lines through the
experimental points are guides to the eye. (lower panel)
Temperature variation of the ordered ferromagnetic moment. The
solid line is the calculated ordered moment supposing a power law.
The inset shows the temperature variation of the spontaneous
orthorhombic strain $(s=2(a-c)/a+c))$. }
\label{abc-r015}%
\end{figure}%

\begin{figure}[tbp] \centering%
\includegraphics[angle=0,width=0.7 \columnwidth]{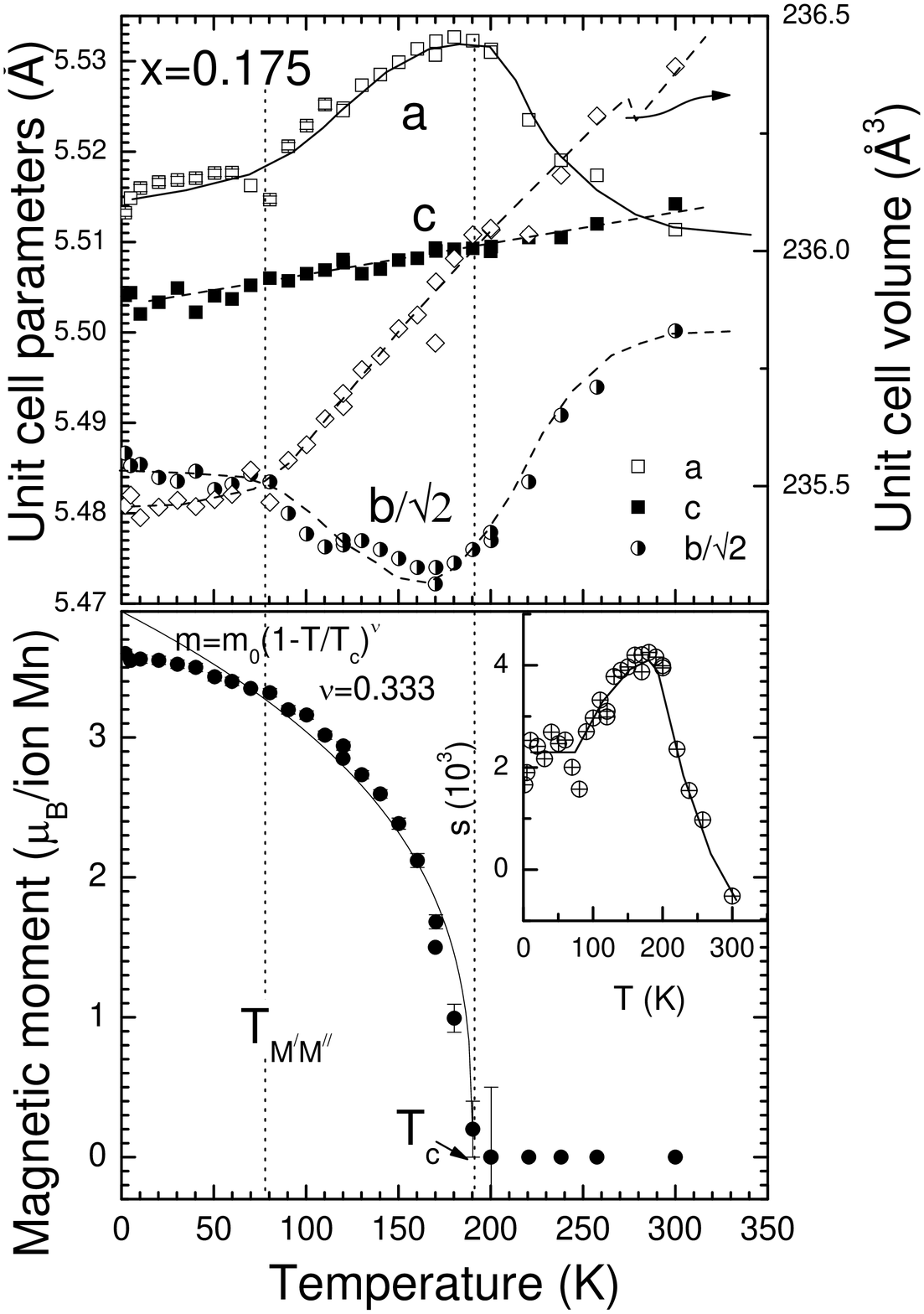}
\caption{ (upper panel) Temperature variation of the lattice
parameters and unit-cell volume for
La$_{0.825}$Ca$_{0.175}$MnO$_{3}$ compound. Lines through the
experimental points are guides to the eye. (lower panel)
Temperature variation of the ordered ferromagnetic moment. The
solid line is the calculated ordered moment supposing a power law.
The inset shows the temperature variation of the spontaneous
orthorhombic strain $(s=2(a-c)/a+c))$.  }
\label{abc-r0175}%
\end{figure}%

The manganite perovskites of interest to this work consist of a 3D
network of corner-sharing BX$_6$ octahedra located at the nodes of
a simple cubic lattice. At the center of the unit cell there is
room for the A cation which will fit perfectly if $d_{\rm
A-X}=d_{\rm B-X}\sqrt{2}$. When $d_{\rm A-X}<d_{\rm B-X}\sqrt{2}$
the BX$_6$ octahedra are tilted to fill the extra interstitial
space (steric effect). The steric effect in
La$_{1-x}$Ca$_x$MnO$_3$ compound leads to a lower symmetry
structure with space group $Pnma$ (in comparison to the aristotype
cubic perovskite $Pm\overline{3}m$ structure) which according to
the Glazer's clasification\cite{glazer} belongs to the $\alpha^-
\beta^+\alpha^-$ tilt system (adopted  by GdFeO$_3$ and CaTiO$_3$
perovskites). The adoption of the GdFeO$_3$ tilting scheme in
La$_{1-x}$Ca$_x$MnO$_3$ is due to the small radius of the La and
Ca, with respect to its surrounding cage. Thus, in order that the
La ion can be accommodated in the structure, the MnO$_6$-octahedra
tilt. Practically MnO$_6$ octahedra rotation is not rigid. Below a
characteristic temperature the La$_{1-x}$Ca$_x$MnO$_3$ compound
for $x<0.23$ displays cooperative JT distortion arising from
ordering of the $e_g$ orbital of the Mn$^{+3}$ ion. This structure
is denoted as\cite{monoclinic} O$^{/}$ and the unit cell satisfies
the inequality $a>c>b/\sqrt{2}$. For this phase, Mn$^{+3}$ $e_g$
orbitals are ordered in such a way so that in neighboring Mn sites
the $|3x^2-r^2\rangle$ and $|3z^2-r^2\rangle$ orbitals are
alternatively occupied, forming a "canted" orbital structure. This
configuration does not change along $b-$axis.

Two mechanisms have been proposed to explain the orbital ordering
in degenerated transition metal orbital states. The first is
related with lattice distortions\cite{kanamori60} and the second
is a direct generalization of the usual superexchange which in the
case of orbital degeneracy is present.\cite{kugel82} The orbital
ordering can be described by using the pseudospin-$1/2$, $T_i$
operator in such a way so that the state $| T^z=\pm 1/2\rangle$
corresponds to the orbitals $|3z^2-r^2\rangle$ and
$|x^2-y^2\rangle$, respectively. An arbitrary distortion can be
described by linear superposition of the states $| T^z=\pm
1/2\rangle$, e.g. $|\theta\rangle=\cos(\theta/2)|
1/2\rangle+\sin(\theta/2)|-1/2\rangle$, where $\theta$ is an angle
in the $(T^z, T^x)$ plane.

Finally, there is a third orthorhombic phase, O*
($c>a>b\sqrt{2}$), related to orbital disordered manganites
($x>0.23$). The O*-structure, is pseudocubic with the Mn-O bond
lengths almost equal, indicating that it lacks a clear static JT
distortion. Both phases exhibit the orthorhombic $Pnma$ symmetry.

In this article we present a synchrotron x-ray and neutron
diffraction study, using carefully prepared stoichiometric
La$_{1-x}$Ca$_x$MnO$_3$ $(0.11\leq x\leq 0.175)$ samples in an
effort to isolate the structural and magnetic features that
characterize the FMI state. Our diffraction results on
stoichiometric samples shows that the crystal structure at FMI
regime is monoclinic $P 2_1/c$. The temperature variation of the
unit cell parameters is non monotonic, defining the $T_{\rm JT}$
and $T_{\rm M^/M^{//}}$ temperatures. In addition, this monoclinic
crystal structure permits two Mn crystallographic sites.  This
crystal structure is different from it of the cation deficient
samples and samples that belong in the ferromagnetic metallic
regime.

\section{Experimental details}

La$_{1-x}$Ca$_{x}$MnO$_{3}$ samples ware prepared by thoroughly
mixing high purity stoichiometric amounts of CaCO$_{3}$,
La$_{2}$O$_{3}$, and MnO$_{2}$. The mixed powders reacted in air
up to 1400$^{\text{o}}$C for several days
with intermediate grinding. Finally the samples were slowly cooled to 25$^{%
{\rm o}}$C. These samples are cation
deficient.\cite{pissas04a,dabrowski99a,roosmalen94} In order to
produce stoichiometric samples additional annealing in He
atmosphere was applied. More specifically, in the final
preparation step the samples were heated under He flow up to
1100$^{\rm o}$C for 48 h followed by rapid cooling to room
temperature.
It is important to stress here that in the doping range examined
in this work the samples fired in air in all preparation steps are
not stoichiometric.
Although these samples are ferromagnetic and insulating for
$x<0.2$, they do not exhibit all the features of the
stoichiometric samples, especially the cooperative Jahn-Teller
(JT) distortion feature for $T<T_{\rm JT}$.

Neutron powder diffraction (NPD) data were collected on the E6 and
E9 neutron diffractometers at the Berlin Neutron Scattering Center
(BENSC), located at the Hahn-Meitner-Institut, Berlin. The neutron
powder diffraction measurements on the E6 diffractometer were
performed as a function of temperature at low angles using (002)
reflection of pyrolytic graphite monochromator, giving a
wavelength $\lambda =2.4$\AA . For high resolution crystal
structure determinations  we measured NPD data using the  E9
diffractometer with a wavelength of $\lambda =1.798$\AA\ ((511)
reflection of a vertically bent Ge monochromator) and collimation
$\alpha _{1}=18^{/}$(in pile collimator), $\alpha
_{2}=20^{/}$(second collimator after monochromator), $\alpha
_{3}=10^{/}$(third collimator after sample). The powdered samples
were contained in a cylindrical vanadium can ($D=8$ mm) mounted in
an ILL orange cryostat.

For high resolution x-ray diffraction data we used the ID31 high
resolution powder diffraction beamline of the European Synchrotron
Radiation Facility (ESRF). For our measurements we select a wave
length $\lambda=0.50019$\AA. Nuclear and magnetic structures were
determined and refined from the NPD data by using the Rietveld
method with the {\sc FULLPROF} suite of programs.\cite{fullprof}
For synchrotron x-ray diffraction data, the GSAS suite was
used.\cite{toby,larson} As peak shape function we used the
gaussians convoluted with axial divergence asymmetry function\cite
{finger94} for neutron data and the constant wave length profile
function no 4 of the GSAS for the x-ray data. DC magnetization
measurements were performed in a superconducting quantum
interference device (SQUID) magnetometer (Quantum Design).
Finally, for the ac susceptibility measurements we used a home
made ac-susceptometer.

\begin{table*}[tbp] \centering%
\caption{La$_{1-x}$Ca$_{x}$MnO$_{3}$ ($x=0.125$) structural
parameters at several temperatures, as determined from Rietveld
refinements based on neutron powder diffraction data of E9
diffractometer ($\lambda=1.79784$\AA ). The monoclinic space group
$P 2_1/c\equiv P\ 1\ 1\ 2_1/a$ (no 14, unique $c$-axis) was used.
In this space group La, Ca O1, O2A and O2B occupy the general $4e$
position $(x,yz)$. Mn1 and Mn2 occupy the $2c$ (1/2,0,0) and $2d$
(1/2,1/2,0) positions respectively. Numbers in parentheses are
statistical errors of the last significant digit.}
\begin{ruledtabular}
\begin{tabular}{ccccccc}
$T$ (K) & & 300 & 198.4 & 121 & 48  & 2 \\
\tableline
$a$ (\AA )     & & 5.5754(1)  & 5.5855(1) & 5.5829(1) & 5.5709(1)&5.5697(1)\\
$b$ (\AA )     & & 7.7423(1)  & 7.7213(1) & 7.7186(1) & 7.7288(1)&7.7296(1)\\
$c$ (\AA )     & & 5.5269(1)  & 5.5253(1) & 5.5234(1) & 5.5206(1)&5.5205(1)\\
$\gamma$ (deg.)& & 90.125(3)  & 90.130(5) & 90.127(4) & 90.106(6)&90.107(5)\\
\tableline
La,Ca &$x$  &  0.0319(4) &  0.0344(3) & 0.0348(6)& 0.0342(3)&0.0341(3) \\
      &$y$  &  0.2501(8) &  0.2487(5) & 0.2477(7)& 0.2479(9)&0.2485(8)\\
      &$z$  &  0.0058(5) &  0.0049(4) & 0.0055(4)& 0.0064(4)& 0.0061(4) \\
      &$B$  &  0.69(4)   &  0.57(3)   & 0.43(3)  & 0.33(3)  &0.37(3)\\
\tableline
Mn1    &$B$  &  0.38(7)   & 0.30(6)    & 0.24(6)  & 0.19(5)  &0.17(6) \\
Mn2    &$B$  &  0.38(7)   & 0.30(6)    & 0.24(6)  & 0.19(5)  &0.17(6) \\
\tableline
O1 &$x$     &  0.4909(6) & 0.4898(5)  & 0.4906(5)& 0.4909(5)&0.4901(5)\\
   &$y$     &  0.244(1)  & 0.247(1)  & 0.246(1) & 0.248(1)&0.247(1)  \\
   &$z$     & -0.0670(6) &-0.0679(4)  &-0.0688(5)&-0.0687(4)&-0.0683(5)\\
   &$B$     &  1.01(6)   & 0.87(5)    & 0.74(5)  & 0.60(4)  &0.62(5)\\
\tableline
O2A&$x$      & 0.213(1) & 0.211(1) & 0.209(1)& 0.215(1)&0.223(1)\\
   &$y$      &-0.0380(9) &-0.0377(8) &-0.0392(7)&-0.0376(8)&-0.0374(9)\\
   &$z$      & 0.784(1) & 0.784(1)  & 0.784(1)& 0.788(1)&0.783(1) \\
   &$B$      & 1.06(5)   & 0.83(4)   & 0.66(4)  & 0.76(4)  &0.73(4)\\
   \tableline
O2B&$x$      &-0.207(1)  &-0.207(1) &-0.208(1)&-0.205(1)&-0.207(1)\\
   &$y$      & 0.4656(9) & 0.4640(7) & 0.4655(7)& 0.4644(8)&0.4641(8)\\
   &$z$      &-0.764(1)  &-0.763(1)  &-0.764(1)&-0.778(1)&-0.766(1)\\
   &$B$      & 1.06(5)   & 0.83(3)   & 0.66(4)  & 0.76(4)  &0.73(4)\\
 \tableline
&$R_{p}$     & 7.0       & 5.9       & 6.4      & 5.7&5.9\\
&$R_{wp}$    & 9.2       & 7.6       & 8.2      & 7.5&7.8 \\
&$R_{B}$     & 6.2       & 5.3       & 5.6      & 4.7&4.8\\
\tableline
Mn1-O1 $\times 2$& &1.928(9) &1.947(8) & 1.942(8) &1.95(1)& 1.948(1)\\
Mn1-O2A$\times 2$& &2.013(7)  & 2.026(6) & 2.035(6) &2.022(8)& 2.014(7)\\
Mn1-O2A$\times 2$& &1.994(8) & 1.986(6) & 1.981(7) &1.971(9)& 1.987(7) \\
Mn2-O1 $\times 2$& &2.014(9)  & 1.986(8)& 1.993(8) &1.98(1)& 1.990(1)\\
Mn2-O2B$\times 2$& &2.102(7)  & 2.111(6) & 2.099(6) &2.082(8)& 2.095(7)\\
Mn2-O2B$\times 2$& &1.884(8)  & 1.881(6) & 1.887(6) &1.911(9)& 1.891(7) \\
\end{tabular}
\end{ruledtabular}\label{table1}
\end{table*}%

\begin{table*}[tbp] \centering%
\caption{La$_{1-x}$Ca$_{x}$MnO$_{3}$ ($x=0.15$) structural
parameters at several temperatures, as determined from Rietveld
refinements based on neutron powder diffraction data of E9
diffractometer ($\lambda=1.79784$\AA ). At $T=320$ K the space
group $Pnma$ (No 62) was used. La, Ca and apical oxygen (O1)
occupy the $4c$, $(x,1/4,z)$ site, Mn the $4b$ $(0,0,1/2)$ site
and the plane oxygen (O2) the general $8d$ site. For other
temperatures the monoclinic space group $P 2_1/c\equiv P\ 1\ 1\
2_1/a$ (no 14, unique $c$-axis) was used. In this space group La,
Ca O1, O2A and  O2B occupy the general $4e$ position $(x,yz)$. Mn1
and Mn2 occupy the $2c$ (1/2,0,0) and $2d$ (1/2,1/2,0) positions
respectively. Numbers in parentheses are statistical errors of the
last significant digit.}
\begin{ruledtabular}
\begin{tabular}{ccccccc}
$T$ (K) & & 320 &250 &  170 & 100  & 2 \\
\tableline
$a$ (\AA )& & 5.5246(1)   & 5.5573(1) & 5.5614(1)&5.5480(1)&5.5300   \\
$b$ (\AA )& & 7.7877(1)   & 7.7387(1) & 7.7266(1)&7.7386(1)&7.7508   \\
$c$ (\AA )& & 5.5219(1)   & 5.5213(1) & 5.5195(1)&5.5159(1)&5.5081   \\
$\gamma$ (deg.)& & 90   & 90.140(1) & 90.134(2)&90.0965(1)&90.0752  \\
\tableline
La,Ca &$x$  &  0.0263(4) &  0.0300(3) & 0.0311(3)& 0.0309(3)&0.0303(2) \\
      &$y$  &  0.25      &  0.2489(5) & 0.2501(7)& 0.2493(6)&0.2454(6)\\
      &$z$  & -0.0083(5) &  0.0060(4) & 0.0060(4)& 0.0058(4)&-0.0065(4) \\
      &$B$  &  0.54(5)   &  0.40(3)   & 0.42(3)  & 0.12(2)  &0.15(3)\\
\tableline
Mn1    &$B$  &  0.36(7)   & 0.25(5)    & 0.19(5)  & 0.19(5)  &0.16(5) \\
Mn2    &$B$  &            & 0.25(5)    & 0.19(5)  & 0.19(5)  &0.16(5) \\
\tableline
O1 &$x$     &  0.4910(8) & 0.4912(4)  & 0.4914(4)& 0.4905(5)&0.4909(4)\\
   &$y$     &  0.25      & 0.2450(8)  & 0.246(1) & 0.2429(8)&0.259(1)  \\
   &$z$     &  0.0684(7) &-0.0668(4)  &-0.0675(1)&-0.0693(4)&-0.0680(5)\\
   &$B$     &  0.94(7)   & 0.69(4)    & 0.58(4)  & 0.45(4)  &0.484(4)\\
\tableline
O2A&$x$      & 0.2817(6) & 0.2180(7) & 0.2188(8)& 0.2212(9)&0.2219(9)\\
   &$y$      & 0.0353(3) &-0.0381(5) &-0.0364(7)&-0.0375(6)&-0.0384(6)\\
   &$z$      & 0.7238(6) & 0.721(1)  & 0.780(1)& 0.780(1)&0.782(1) \\
   &$B$      & 0.88(5)   & 0.63(3)   & 0.58(3)  & 0.40(3)  &0.32(3)\\
\tableline
O2B&$x$      &           &-0.2075(8) &-0.2059(9)&-0.206(1)&-0.2085(0)\\
   &$y$      &           & 0.4659(6) & 0.4639(7)& 0.4657(7)&0.4666(6)\\
   &$z$      &           &-0.766(1)  &-0.768(1)&-0.768(1)&-0.768(1)\\
   &$B$      &           & 0.63(3)   & 0.58(3)  & 0.40(3)  &0.32(3)\\
 \tableline
&$R_{p}$     & 6.9       & 5.1       & 6.2      & 6.0&6.1\\
&$R_{wp}$    & 9.0       & 6.8       & 7.9      & 8.0&7.9 \\
&$R_{B}$     & 4.3       & 4.1       & 5.6      & 3.7&3.6\\
\tableline
Mn1-O1 $\times 2$& &1.9839(8) &1.932(6) & 1.941(8) &1.919(6)& 1.973(7)\\
Mn1-O2A$\times 2$& &2.006(3)  & 1.999(5) & 1.996(6) &1.986(6)& 1.973(6)\\
Mn1-O2A$\times 2$& &1.9634(3) & 1.992(5) & 1.991(6) &1.995(7)& 2.002(7) \\
Mn2-O1 $\times 2$& &          & 2.007(6)& 1.994(8) &2.026(6)& 1.975(7)\\
Mn2-O2B$\times 2$& &          & 2.092(5) & 2.094(6) &2.086(6)& 2.071(6)\\
Mn2-O2B$\times 2$& &          & 1.886(2) & 1.893(6) &1.891(7)& 1.893(6) \\
\end{tabular}
\end{ruledtabular}\label{table3}
\end{table*}%

\begin{table*}[tbp] \centering%
\caption{La$_{1-x}$Ca$_{x}$MnO$_{3}$ ($x=0.175$) structural
parameters at several temperatures, as determined from Rietveld
refinements based on neutron powder diffraction data of E9
diffractometer ($\lambda=1.79784$\AA ). At $T=300$ K the space
group $Pnma$ (No 62) was used. La, Ca and apical oxygen (O1)
occupy the $4c$, $(x,1/4,z)$ site, Mn the $4b$ $(0,0,1/2)$ site
and the plane oxygen (O2) the general $8d$ site. For other
temperatures the monoclinic space group $P 2_1/c\equiv P\ 1\ 1\
2_1/a$ (no 14, unique $c$-axis) was used. In this space group La,
Ca O1, O2A and  O2B occupy the general $4e$ position $(x,yz)$. Mn1
and Mn2 occupy the $2c$ (1/2,0,0) and $2d$ (1/2,1/2,0) positions
respectively. Numbers in parentheses are statistical errors of the
last significant digit.}
\begin{ruledtabular}
\begin{tabular}{cccccccc}
$T$ (K) &   & 300       & 200      & 170        & 120       & 70       & 2      \\
\tableline
$a$ (\AA )& & 5.5113(1) &5.5275(1) & 5.5323(1)  & 5.5242(1)& 5.5165(1)&5.5152(1)\\
$b$ (\AA )& & 7.7778(1) &7.7449(2) & 7.7414(1)  & 7.7453(1) & 7.7553(1)&7.7568(1)\\
$c$ (\AA )& & 5.5142(1) &5.5114(1) & 5.5104(1)  & 5.5080(1) & 5.5050(1)&5.5042(1)\\
$\gamma$ (deg.)& 90.0  &  &90.094(6) & 90.09(1)  & 90.063(5)  &90.059(4)  &90.056(4)  \\
\tableline
La,Ca &$x$  &  0.0252(4)&0.0283(3)& 0.0292(5)&  0.0290(3) & 0.0280(3)& 0.0278(3) \\
      &$y$  & 0.25     &0.248(1)& 0.244(1)&  0.247(1) & 0.249(1)& 0.248(1) \\
      &$z$  & -0.008(1) &0.0063(5)&0.0087(7)&  0.0074(5) & 0.0070(6)&0.0077(6) \\
      &$B$  &  0.57(4)  &0.44(3)   & 0.70(4)  &  0.37(3)   & 0.28(3)  & 0.28(3)\\
\tableline
Mn1    &$B$  &  0.37(7)  &0.1(1)   &0.34(8)   & 0.08(6)    & 0.11(6)  & 0.13(6)\\
Mn2    &$B$  &           &0.1(1)   &0.34(8)   & 0.08(6)    & 0.11(6)  & 0.13(65)\\
\tableline
O1 &$x$     &  0.4927(8)&0.4914(6)&0.4910(8) & 0.4927(6) & 0.4913(6)& 0.4913(5)\\
   &$y$     &  0.25     &0.247(1)&0.246(1) & 0.248(1) & 0.250(1)& 0.251(1)\\
   &$z$     &  0.0680(8)&-0.0669(5)&-0.0646(7)&-0.0671(5) &-0.0665(6)&-0.0673(3)\\
   &$B$     &  0.86(7)  &0.65(5)  &0.77(6)   & 0.46(5)   & 0.38(5)  & 0.37(5) \\
\tableline
O2A&$x$      & 0.2792(7)&0.221(1)&0.221(2) & 0.224(1) & 0.221(1)&0.221(1)\\
   &$y$      & 0.0344(4)&-0.035(1)&-0.0360(9)&-0.0369(8) &-0.033(1)&-0.033(1)\\
   &$z$      & 0.7240(7)&0.781(1)&0.771(2) & 0.770(1) & 0.781(1)& 0.778(2)\\
   &$B$      & 0.7224(7)&0.62(3)  &1.04(5)   & 0.64(4)   & 0.50(3)  & 0.56(4) \\
\tableline
O2A&$x$      &          &-0.209(1)&-0.209(2) &-0.208(1) &-0.213(1)&-0.214(1)\\
   &$y$      &          &0.464(1)&0.4676(9) & 0.4666(9) & 0.462(1)& 0.462(1)\\
   &$z$      &          &-0.766(1)&-0.771(2) &-0.780(1) &-0.770(1)&-0.774(2)\\
   &$B$      &          &0.62(4)  &1.04(5)   & 0.64(4)   & 0.50(3)  & 0.56(4) \\
\tableline
&$R_{p}$     & 6.0      &5.4      & 6.8      & 5.8       & 6.1       & 6.2\\
&$R_{wp}$    & 8.0      &7.0      & 9.4      & 7.6       & 8.0       & 8.0 \\
&$R_{B}$     & 3.3      &3.6      & 5.6      & 4.2       & 3.8       & 3.8\\
Mn1-O1$\times 2$&&1.9809(1) & 1.949(1)& 1.93(1)& 1.95(1) &1.979(1) & 1.98(1) \\
Mn1-O2A$\times 2$&&1.996(4) & 1.973(8) & 2.00(1) & 2.000(8)  &1.968(8) & 1.980(9)\\
Mn1-O2A$\times 2$&&1.973(4) & 1.995(8) & 1.95(1) & 1.959(8)  &1.989(9)  & 1.97(1) \\
Mn2-O1$\times 2$&&          & 1.993(1)& 1.99(1)& 1.98(1) &1.96(1) & 1.96(1) \\
Mn2-O2B$\times 2$&&         & 2.074(8) & 2.04(1) & 2.029(8)  &2.043(8) & 2.028(9)\\
Mn3-O2B$\times 2$&&         & 1.891(8) & 1.92(1) & 1.944(9)  &1.922(8)  & 1.93(1) \\

\end{tabular}\end{ruledtabular}\label{table4}
\end{table*}%


\section{Crystal structure refinements}
Our structural and magnetization results on stoichiometric
La$_{1-x}$Ca$_{x}$MnO$_{3}$, with an emphasis on the FMI regime,
are summarized in Fig. \ref{diagram} in the form of a phase
diagram. The particular phase diagram is in good agreement with
the one estimated by Biotteau et al., \cite{biotteau01} using
single crystal samples. The $T_{\rm JT}$ curve is defined at the
temperature where the magnetization curves display a jump-like or
slope-change behavior and the unit cell parameter $a$ ($b$)
increases (decreases) abruptly on cooling. The particular behavior
of the unit cell constants resemble the situation observed in
LaMnO$_3$ compound on cooling through the $T_{\rm JT}$
transition.\cite{carvajal98} Depending on $x$, for $T<T_{\rm JT}$
a symmetry change from $Pnma$ to $P 2_1/c$ occurs. The $T_{\rm
C}(x)$ and $T_{\rm N}(x)$ lines concern the Curie and Neel
temperatures as were estimated both from neutron and magnetization
data. The line $T_{\rm M^/M^{//}}(x)$ is defined at the
temperature where the $a(T)$ and $b(T)$ curves display an abrupt
slope change, upon heating from $T=0$. Based on the transition
lines mentioned above the phase diagram of the
La$_{1-x}$Ca$_{x}$MnO$_{3}$ compound depicted in Fig.
\ref{diagram} displays two distinct regions as the ground state is
concerned. In the first region a canted antiferromagnetic (CAF)
orthorhombic structure is present. In the second one, depending on
$x$, a ferromagnetic insulating (FMI) monoclinic and orthorhombic
structures have been observed. The change from CAF to FMI ground
states occurs in a narrow concentration regime $0.12<x\lessapprox
0.14$, where in neutron diffraction patterns (vide infra) a
ferromagnetic transition ($T_{\rm C}$) is observed firstly and
after an antiferromagnetic transition ($T_{\rm N}$) have been
observed upon cooling.  Our high resolution synchrotron x-ray
diffraction patterns revealed that the true symmetry of the
samples in the boundary of the CAF and FMI regimes and partially
in the FMI regime is monoclinic with very small monoclinic angle
($\gamma=0.04-0.1^{\rm o}$). This regime is denoted by symbol
M$^/$ in the phase diagram. The temperature variation of the unit
cell parameters in the FMI regime revealed an additional
structural transformation which occurs upon heating from $T=0$ at
temperature $T_{\rm M^/M^{//}}$. The magnitude of the unit cell
changes at the $T_{\rm M^/M^{//}}$ reduces when the $x$ is in the
CAF-FMI(M) and FMI(M$^/$)-FMI(O$^/$) phase boundaries. In this
phase diagram the ferromagnetic to paramagnetic ($T_{\rm C}$), JT
($T_{\rm JT}$) and $T_{\rm M^/M^{//}}$ transition lines are all
estimated from neutron and magnetic measurements.
\begin{figure*}[tbp] \centering%
\includegraphics[angle=0,width=5in]{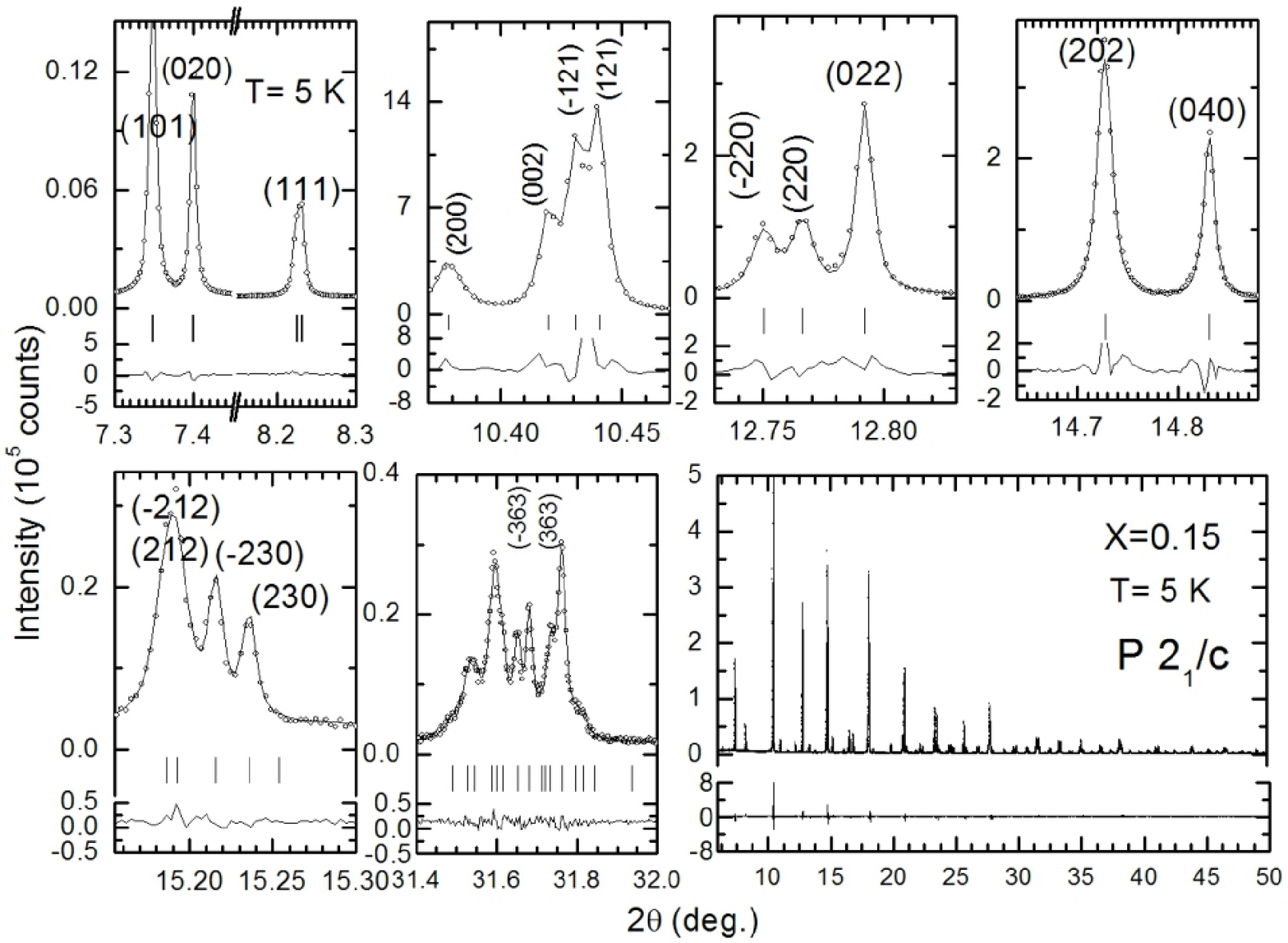}
\caption{ Rietveld plot for La$_{0.85}$Ca$_{0.15}$MnO$_{3}$ at
$T=5$ K using the high resolution x-ray diffraction data from ID31
beam line of the ESRF. Show are details of the diffraction peaks
where the characteristic peak splitting compatible with $P 2_1/c$
space group is expected.}
\label{esrf}%
\end{figure*}%

\begin{figure}[tbp] \centering%
\includegraphics[angle=0,width=0.8 \columnwidth]{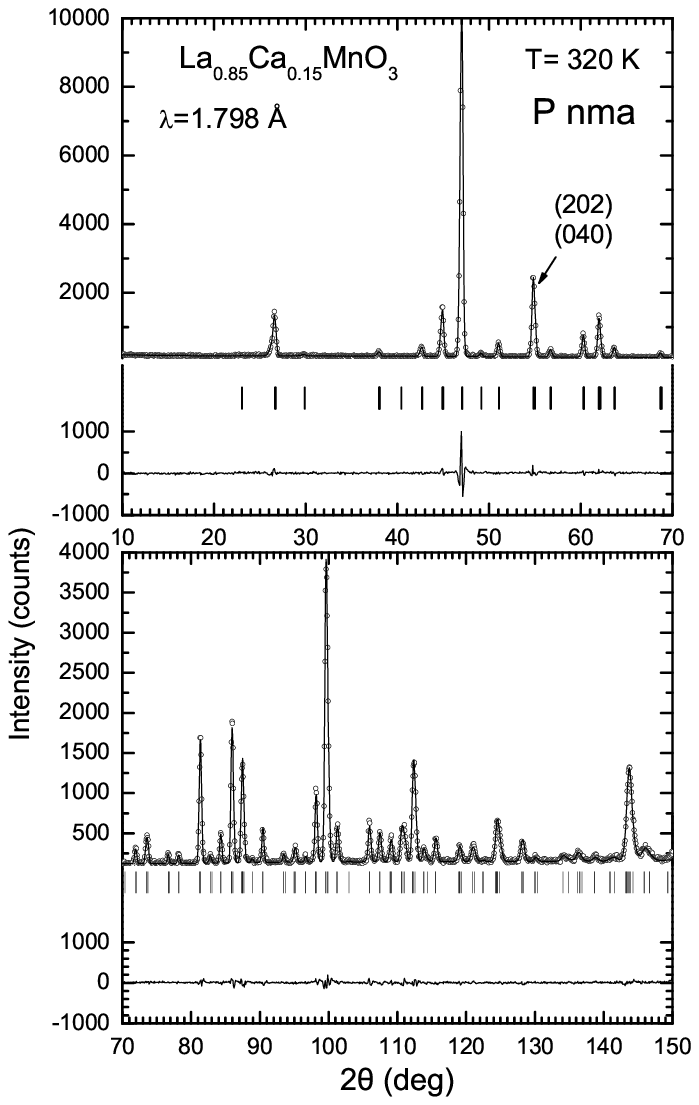}
\caption{ Rietveld plot for La$_{0.85}$Ca$_{0.15}$MnO$_{3}$ at
$T=320$ K. The observed data points are indicated by open circles,
while the calculated and difference patterns are shown by solid
lines. The positions of the reflections are indicated by vertical
lines below the patterns.}
\label{rietveld-320-r015}%
\end{figure}%

\begin{figure}[tbp] \centering%
\includegraphics[angle=0,width=0.8 \columnwidth]{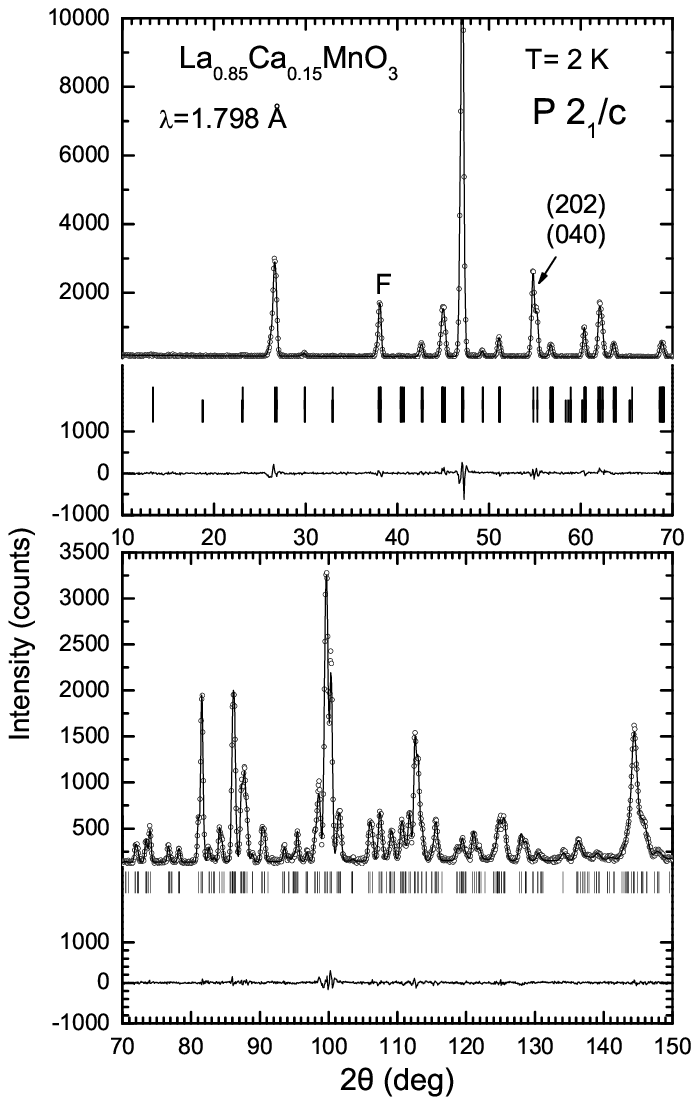}
\caption{ Rietveld plot for La$_{0.85}$Ca$_{0.15}$MnO$_{3}$ at
$T=2$ K. The observed data points are indicated by open circles,
while the calculated and difference patterns are shown by solid
lines. The positions of the reflections are indicated by vertical
lines below the patterns.
}
\label{rietveld-5-r015}%
\end{figure}%
 Figures \ref{abc-r013}, \ref{abc-r015} and \ref{abc-r0175}
(upper panels) show the temperature variation of the unit cell
parameter for $x=0.13, 0.15$ and 0.175 samples using the pseudo
orthorhombic description of the unit cell.
For $T<T_{\rm JT}$, the unit cell parameters follow the inequality
$a>c>b/\sqrt{2}$, a fact implying that in this pseudo orthorhombic
description the ${\rm O^/}$ structure is realized in the whole
temperature regime studied in this work. For all the samples there
is a temperature region where the unit cell parameters change
abruptly ($a$ increases, $b/\sqrt{2}$ decreases and $c$ decreases
slightly) resembling the situation observed\cite{carvajal98} in
the LaMnO$_3$  compound due to the orbital ordering state. As it
is mentioned above, we define the transition temperature $T_{\rm
JT}$ as the point where an extreme in the $da/dT$ curve occurs, in
the high temperature regime of the data. This particular
assignment agrees with the magnetic measurements (vide infra). The
orthorhombic strain parameter $s=2(a-c)/(a+c)$ (see Figs.
\ref{abc-r013}, \ref{abc-r015} and \ref{abc-r0175}, lower panels)
becomes maximum at the ferromagnetic transition temperature.
Furthermore, for $T<T_c$ $s$ decreases during cooling and finally
for $T<T_{\rm M^/M^{//}} \sim 60$ K (for the $x=0.13$ sample) it
becomes almost temperature independent. The particular behavior
reveal that the FMI state does not concern only the spin degrees
of freedom but is also related with the lattice and orbital ones.
The temperature variation of the unit cell volume $V(T)$ exhibits
a discontinuous change at $T_{\rm JT}$ implying that the JT
distortion is first order. As $x$ increases the volume jump is
difficult to be identified, on the basis of the available
diffractometer resolution. It is very likely that the JT
structural transition becomes second order for $x\ge 0.175$. Above
and below  $T_{\rm JT}$ the unit cell volume decreases with
temperature nearly linearly, showing a small slope change at
$T_{\rm C}$. Finally, at $T_{\rm M^/M^{//}}$ $V(T)$ displays a
pronounced slope change for all samples.

Let us present now the high resolution synchrotron x-ray data. The
samples for $0.125\leq x\leq 0.175$ for $T<T_{\rm JT}$ exhibited
additional reflections that can not be accounted for the $Pnma$
space group. More specifically, we observed splitting of the
$(hk0)$, and $(hkl)$ diffraction peaks not allowed in the $Pnma$
space group. This splitting is compatible with the $P 1 1
2_1/a\equiv P 2_1/c$ (no 14, unique $c$-axis) space group. Fig.
\ref{esrf} depicts the high resolution synchrotron x-ray
diffraction data for the La$_{0.85}$Ca$_{0.15}$MnO$_3$ sample at
$T=5$ K, as an example. One can see very clearly the splitting of
the diffraction peaks with non zero $h$ and $k$. We obtained
similar data also for the other samples $x=0.125-0.175$. In this
monoclinic distortion $a$ glide-plane element is preserved, while
the $n$ and $m$ elements are lost. The loss of the mirror plane
may be signaling a change of the JT order from C type orbital
ordering (that is followed by LaMnO$_3$) to a new one. We note
that similar monoclinic distortion have been detected in
La$_{1-x}$Sr$_x$MnO$_3$ ($x=0.11-0.5$) compounds.\cite{cox01}
Based on this critical information about the symmetry of the
structure we refine the neutron data using the $P 2_1/c$ space
group.
The mirror plane at $y=1/4$ being present for $Pnma$ space group
is now absent, resulting in two Mn sites along the $b$ axis, the
first at $(1/2,0,0)$ and the second at $(1/2,1/2,0)$. Similarly,
instead of one plane oxygen site (O2, $P nma$ symmetry) there are
two now in the $P 2_1/c$ symmetry. Thus, in this new symmetry two
crystallographically independent Mn sites produce a layer type
arrangement of two different MnO$_{6}$ octahedra along the $b$
axis.
The refinement strategy employed in this work was to first
optimize the scale factor, background, unit cell parameters and
zero-shift errors. Subsequently, we refined the atomic positional
and isotropic thermal parameters. In order to reduce the number of
parameters we decided to refine the thermal parameters for all the
sites isotropically and we kept equal the thermal parameters of
the Mn and O2 sites.  Figures \ref{rietveld-320-r015} and
\ref{rietveld-5-r015} show two representative Rietveld plots of
neutron data for La$_{0.85}$Ca$_{0.15}$MnO$_{3}$ sample at 320 K
and 2 K, respectively. The corresponding structural parameters for
all the studied samples are reported in Tables
\ref{table1}-\ref{table4}.

\begin{figure}[tbp] \centering%
\includegraphics[angle=0,width=3in]{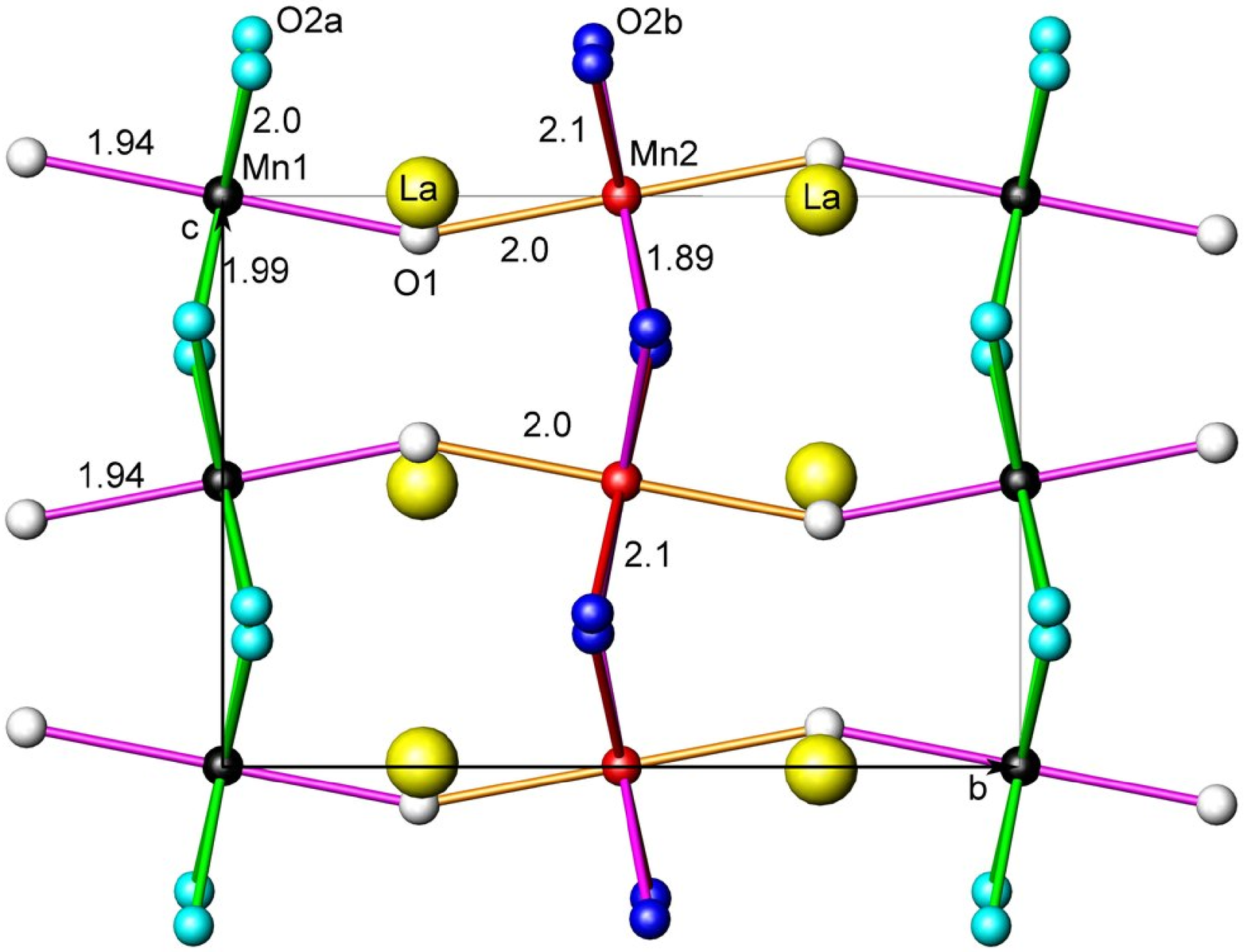}
\includegraphics[angle=0,width=3in]{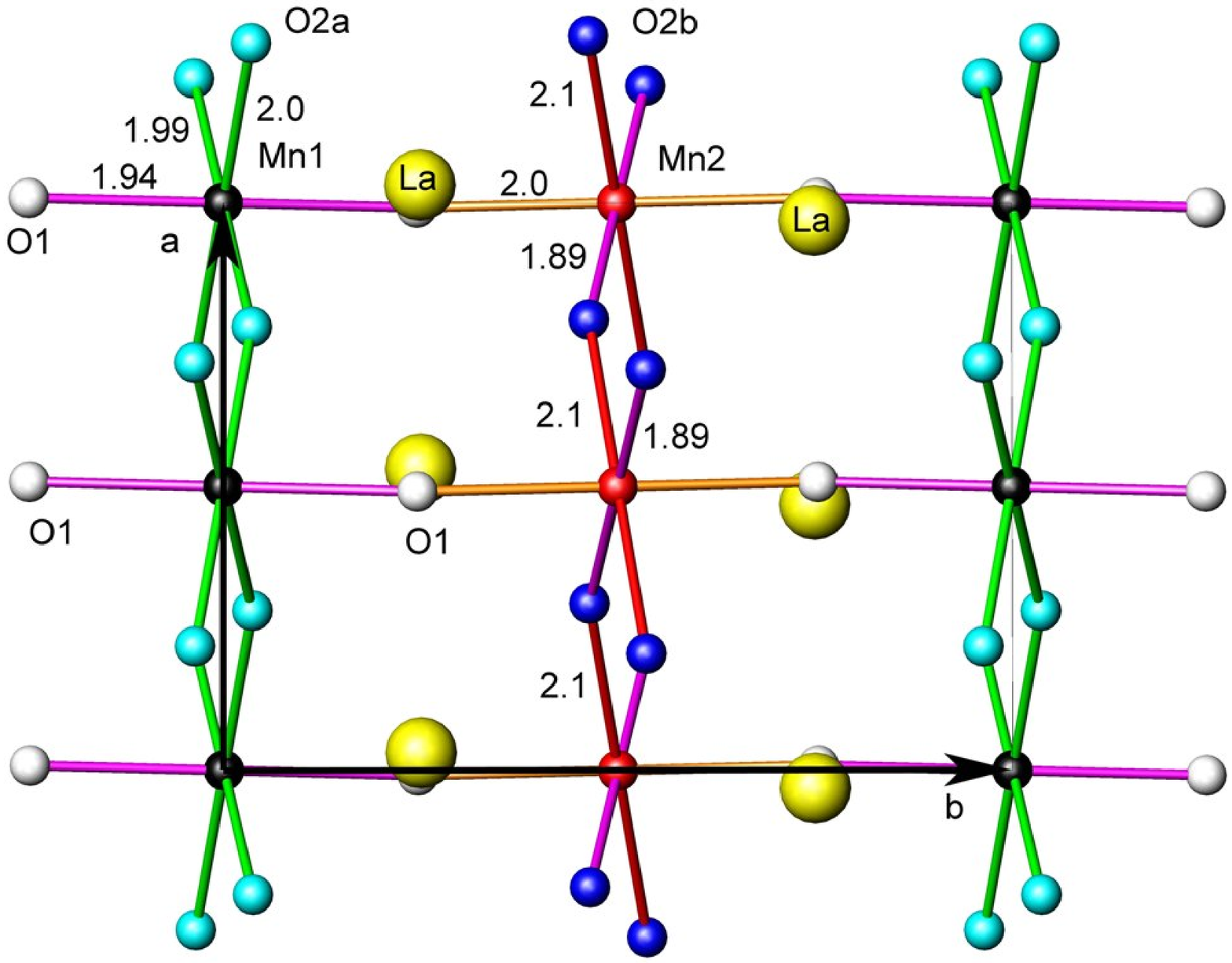}
\caption{Crystal structure of La$_{0.85}$Ca$_{0.15}$MnO$_{3}$
compound at $T=170$ K. The upper and lower panels show the
projections along the $a$ and $c$ axis, respectively. The large
hatched, small filled and intermediate open spheres represent La,
Mn and O ions respectively. The Mn1, Mn2 sites are located at
$y=0$ and $y=1/2$ planes respectively ($b-$axis is the largest
one). The sort Mn1-O, Mn2-O bonds denoted by grey rods are
directed along and perpendicular to the $b$-axis, respectively.}
\label{structure}%
\end{figure}

Fig. \ref{structure} shows projections of the monoclinic structure
of the La$_{0.85}$Ca$_{0.15}$MnO$_{3}$ compound for $T=170$ K on
the $bc$ and $ab$ planes, respectively. All the samples show
qualitatively the same behavior, as far as the variation of the
structural parameters with temperature is concerned. Nevertheless,
there are significant quantitative differences related to the
influence of the hole doping through Ca substitution.

The Bragg peaks are quite sharp for $T>T_{\rm C}(x)$ but for
$T<T_{\rm JT}$ they develop a pronounced selective peak
broadening. In order to account for the selective peak broadening
the Stephens\cite{stephens99} microstrain model was used. As our
Rietveld refinement results reveal the microstrain parameters
follow the temperature variation of the unit cell parameters. The
most significant broadening is related to $S_{202}$ and $S_{400}$
parameters. The $S_{202}$ strain parameter describes a correlated
type random strain between $a$ and $c$ crystallographic directions
and it is reasonable to be attributed to some kind of local static
orbital ordering correlations. The average structure revealed a
cooperative JT distortion which increases as  temperature
decreases. Parallel to this behavior, there is an antagonistic
interaction which produces correlated fluctuation of the cell
constants. These fluctuations are directly connected with the
development of the ferromagnetic long range order and temperature
variation of the unit cell parameters. The microstrain parameters
take the maximum value at $T_{\rm C}$. Most importantly the
microstrain parameters are reduced about 30\% at $T_{\rm
M^/M^{//}}$ and below this temperature remain nearly temperature
independent. Consequently, our structural data reveal a complex
magneto-elastic behavior for $T<T_{\rm C}$, that characterizes the
FMI regime of the La$_{1-x}$Ca$_{x}$MnO$_{3}$ compound.

Anisotropic broadening is usually connected with the presence of
correlated structural defects, due to the preparation method, or
is inherent to  structural or magnetic transformations. Since we
observe temperature dependance in the microstrain parameters, the
case of the structural defects caused by the preparation method
should be ruled out, because in this case a temperature
independent behavior is expected.  It is natural one to ask what
is the physical origin of the particular temperature variation of
the microstrain parameters. Essentially, the anisotropic peak
broadening concerns the lattice planes $(h00)$ and $(h0l)$,
meaning that there is some modulation along the $a$-axis and ${\bf
a}+{\bf c}$ -direction.
Our findings may be related with recent theoretical studies where
the important role of the elastic energies (strain) in determining
the ground state of the system has been pointed
out.\cite{calderon03,ahn04} Furthermore, the anisotropic
broadening  may originate from grains which contain different
structural twin domains corroborated by the successive
magneto-structural transformation occurring upon cooling.
Finally, one cannot disregard the case where phase separation is
present especially near the CAF-FMI phase boundary. In such a case
several phases with nearly similar parameters occur producing
indirectly a selective peak broadening, which is accounted for by
microstrain formalism.

\section{Bond lengths}
\begin{figure}[tbp] \centering%
\includegraphics[angle=0,width=0.8 \columnwidth]{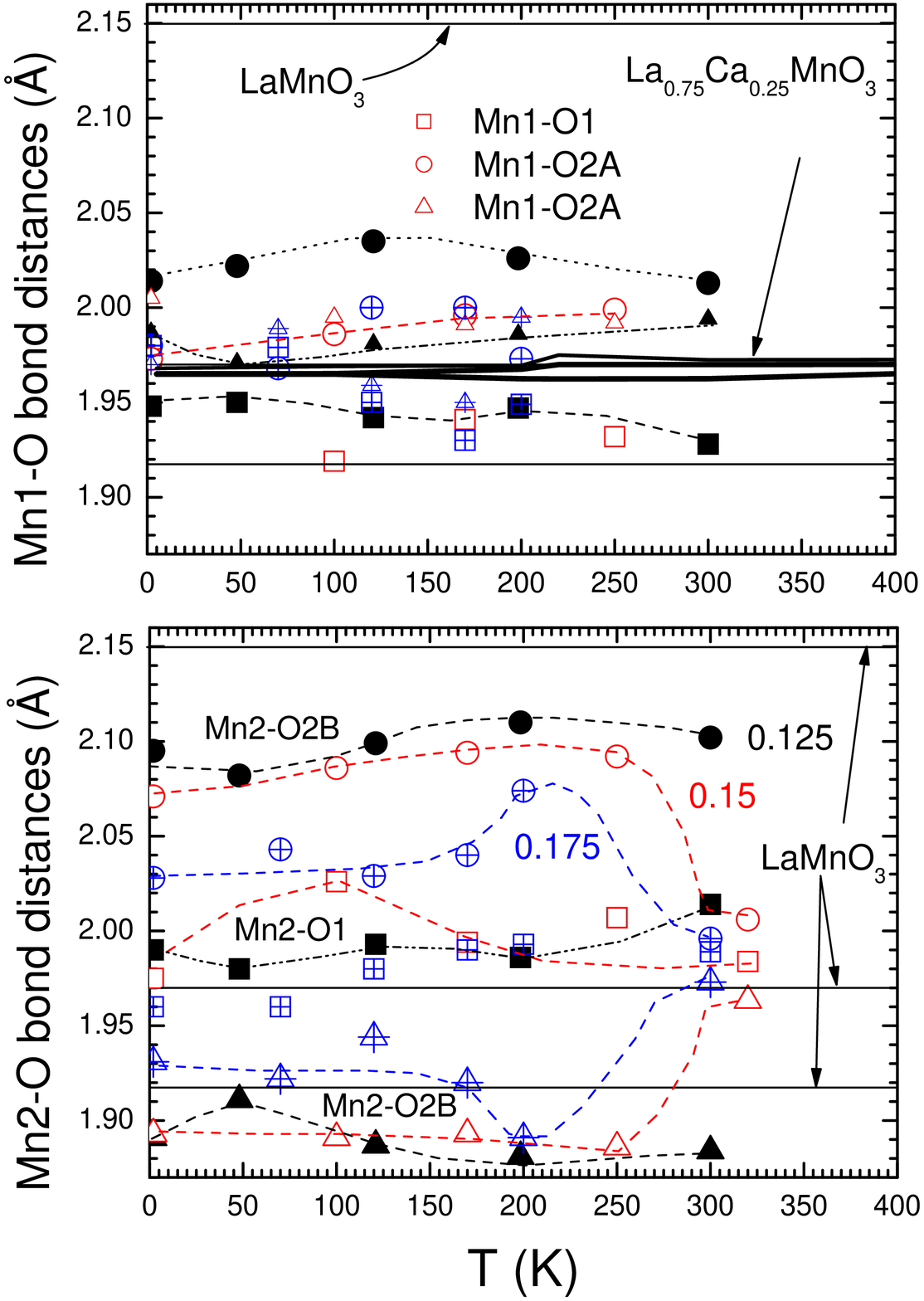}
\caption{ Mn-O$_i$ ($i=1,2$) bond distances as a function of
temperature for $x=0.125, 0.15$ and 0.175 samples. The upper panel
shows the Mn1-O bonds and lower panel the Mn2-O bonds (Mn1-O1,
Mn2-O1 are the apical bonds and Mn1-O2A, Mn2-O2B are the in-plane
bonds). Lines through the experimental points are guides to the
eye. The solid, open and "crossed" symbols correspond to
$x=0.125$, 0.15 and 0.175 respectively. The data for the LaMnO$_3$
(thin lines) and La$_{0.75}$Ca$_{0.25}$MnO$_3$ (thick lines)
samples are also included for a direct comparison.}
\label{bond}%
\end{figure}%

In Fig. \ref{bond} are depicted the temperature variation of the
Mn-O bond lengths for  $x=0.125, 0.15$ and 0.175 samples, deduced
from neutron diffraction data. For $T<T_{\rm JT}$ the crystal
structure of the La$_{1-x}$Ca$_{x}$MnO$_{3}$ in the FMI regime
based on the $P 2_1/c$ space group have two Mn sites.
As the Ca content increases the long Mn2-O bond length (which has
been attributed to the $d_{z^2}$ orbital) decreases. The short
Mn2-O bonds are approximately at the same level as in the
LaMnO$_3$ case. Concerning the intermediate Mn2-O bond, its
magnitude is above the corresponding one of LaMnO$_3$. Is is
interesting to note that the long and short bonds follow the
temperature variation of the $a$ and $c$ cell parameters or in
other words these bonds define the temperature variation of the
unit cell parameters. Based on the similarities of the Mn2-O bond
lengths with those of LaMnO$_3$, it is reasonable to attribute the
Mn2 site to a Mn$^{+3}$ like site in the
La$_{1-x}$Ca$_{x}$MnO$_{3}$  ($0.125\le x \le 0.175$) compound.

Let us turn now to the Mn1-O bond lengths. As their magnitude is
concerned these bond lengths, are more close to the bond lengths
of MnO, observed in the FMM region of the phase diagram. In Fig.
\ref{bond} (upper panel) with bold lines, the Mn-O bond lengths
observed in La$_{0.75}$Ca$_{0.25}$MnO$_3$ compound are depicted.
One can see that as $x$ increases the Mn1-O bond lengths approach
these lines. It is plausible to attribute the Mn1 site to a mixed
valence Mn like site as observed in the
La$_{0.75}$Ca$_{25}$MnO$_3$ compound. As $x$ reduces towards the
CAF regime the Mn1-O bond lengths become non degenerated and move
away from those of La$_{0.75}$Ca$_{0.25}$MnO$_3$
compound.\cite{radaelli97a}

Except for the metrical variation of the bond lengths there is an
orientation difference, concerning the short Mn$i$-O bonds
($i=1,2$), between the two Mn-sites. In the Mn2-site (which
resembles the LaMnO$_3$ case) the short bond lies in the $ac$
plane, contrary to the Mn1-site, where the short Mn1O-bond is
directed, approximately, along the $b$-axis.
According to the authors' opinion this is the
 key point to which one can be based in order
to explain the FMI phase. It is important to note the fundamental
difference between the conventional JT orthorhombic O$^/$ and the
monoclinic M$^/$ structures. The latter cannot be the usual C-type
antiferrodistortive arrangement found in the O$^/$ structure in
LaMnO$_3$ and CAF part of the phase diagram. As proposed by Cox et
al\cite{cox01} in the case of La$_{1-x}$Sr$_x$MnO$_3$, the loss of
the mirror plane would be consistent with a change in JT order
from C-type to G-type at which there is antiferrodistortive
coupling of MnO$_6$ octahedra along the $b$ direction instead of
parallel coupling (see Fig. \ref{structure}).

\section{Magnetic measurements and magnetic structure}

\begin{figure}[tbp] \centering%
\includegraphics[angle=0,width=0.85 \columnwidth]{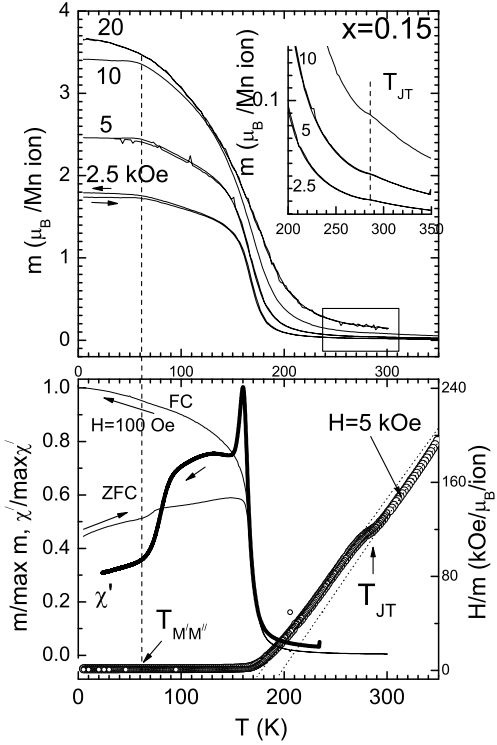}
\caption{ (upper panel) Temperature variation of the bulk magnetic
moment per ion Mn for La$_{0.85}$Ca$_{0.15}$MnO$_{3}$. The data
were collected under the indicated magnetic fields using the
standard ZFC-FC procedure. The inset shows the details of the
measurements at the $T_{\rm JT}$ region. (lower panel) Temperature
variation of the inverse dc magnetic moment, normalized real part
of the fundamental ac-susceptibility (bold line) and normalized
ZFC and FC bulk magnetic moment measured under a dc field 100 Oe
(thin line). The dot lines corresponds to the Curie-Weiss law used
to fit the data above and below $T_{\rm JT}$. The vertical dashed
line indicates the $T_{\rm O^/O^{//}}$. }
\label{moment15}%
\end{figure}

\begin{figure}[tbp] \centering%
\includegraphics[angle=0,width=0.85 \columnwidth]{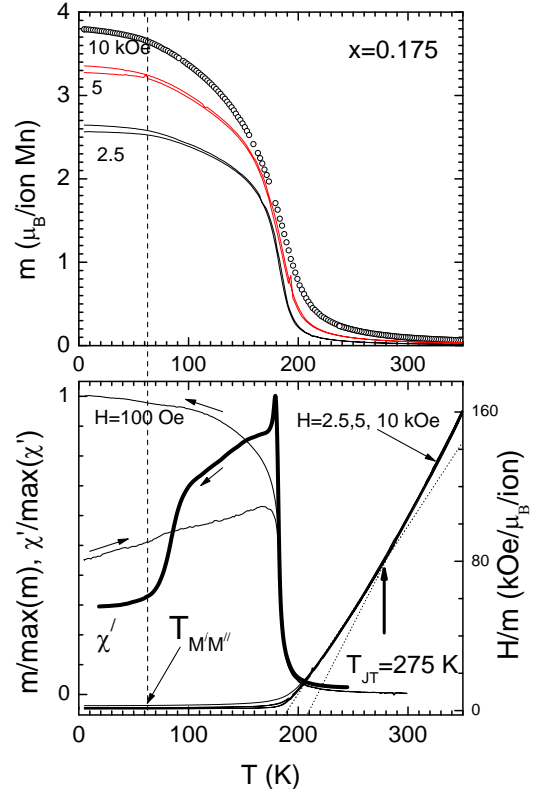}
\caption{ (upper panel) Temperature variation of the bulk magnetic
moment per ion Mn for La$_{0.825}$Ca$_{0.175}$MnO$_{3}$. The data
were collected under the indicated magnetic fields using the
standard ZFC-FC procedure. (lower panel) Temperature variation of
$H/m$, normalized real part of the fundamental ac-susceptibility
(bold line) and normalized ZFC and FC bulk magnetic moment
measured under a dc magnetic field 100 Oe (thin line). The dot
lines corresponds to the Curie-Weiss law used to fit the data
above and below $T_{\rm JT}$.  The vertical dashed line indicates
the $T_{\rm M^/M^{//}}$.}
\label{moment175}%
\end{figure}

Fig.~\ref{moment15} and Fig.~\ref{moment175} (upper panels) show
the temperature variation of the magnetic moment per Mn ion for
the samples $x=0.15$ and 0.175, respectively, for several dc
magnetic fields. We obtained similar magnetic measurements data
also for $x=0.125,0.13$ and 0.14 samples. In the lower panels of
Figs.~\ref{moment15}, and \ref{moment175} we plot the temperature
dependance of the normalized real part of the ac susceptibility,
and the low field dc magnetic moment (left axis). In addition, we
plot the temperature variation of the inverse dc susceptibility
($H/m$) (right axis) measured under a dc-field $H=5$ kOe. The
abrupt increasing of the magnetic signal in the dc and ac
measurements at $T_{\rm C}=168$ K and $T_{\rm C}=187$ K for
$x=0.15$ and $x=0.175$ respectively marks a transition from a
paramagnetic to ferromagnetic state, in agreement with neutron
data (vide infra). The sharp peak at $T_{\rm C}$ in the $\chi'$ is
related with the Hopkison effect, as we discussed in Ref.
\onlinecite{pissas04a}. Upon heating from $T=0$, $\chi'(T)$
slightly increases up to $T\approx T_{\rm M^/M^{//}}$. For
$T>T_{\rm M^/M^{//}}$ $\chi'(T)$ increases in a step wise fashion
as the dashed vertical lines in Figs. \ref{moment15} and
\ref{moment175} show. This peculiarity reveals the magnetic
signature of the structural change that occurs at $T_{\rm
M^/M^{//}}$. The particular anomaly is present and in the dc
magnetization measurements as it is seen for example, in the upper
panel of the Fig. \ref{moment15} where a step like slope change
occurs at $T_{\rm M^/M^{//}}$.

Furthermore, a close inspection of the $m(T)$ curves at the
paramagnetic regime reveals an additional anomaly which is related
with JT distortion observed in the crystal structure data (see
inset of Fig. \ref{moment15}). This anomaly is more pronounced for
$x=0.11$ sample, its signature reduces as $x$ increases and
finally becomes hard to discern for the samples with $x\ge 0.175$.
The anomaly in the magnetic measurements at $T_{\rm JT}$ is seen
more clearly in the inverse susceptibility data (see  lower panels
of Fig.~\ref{moment15} and Fig.~\ref{moment175}). Comparing these
data one can clearly see that the jump of $H/m$ at $T_{\rm JT}$ is
transformed to a slope change as $x$ increases. The particular
change at $T_{\rm JT}$ for the $x> 0.175$ samples may mark a
change of JT transition from first to second order.

 Above and below $T_{\rm JT}$, $\chi^{-1}$ can be fitted by a
straight line implying a Curie-Weiss law
($\chi^{-1}=(T-\Theta)/C)$. The Curie constant is given by
$C=(N/V_c)g^2\mu_B^2S(S+1)/3k_B$, where $N$ is the number of
magnetic ions inside the unit cell, $V_c$ is the unit cell volume,
$g\approx 2$ ($S=J, L=0)$ is the Land\'{e} factor, $\mu_B$ is Bohr
magneton, $S$ the spin of the ion, $k_{\rm B}$ is Boltzmamn's
constant and $\Theta$ is the Weiss constant. Since $\chi^{-1}$ is
nearly parallel above and below $T_{\rm JT}$ it is reasonable to
attribute this change to $\Theta$. Therefore, since $\Theta\propto
J$ (where $J$ is the exchange coupling) the abrupt change in the
$\chi^{-1}$ may be related to a discontinuous change of $J$ above
and below $T_{\rm JT}$. The O$^/$ phase has larger exchange
constants in comparison to the M$^{/}$ one. The intersection of
$H/m$ with the temperature-axis, for $T<T_{\rm JT}$ gives a
$\Theta=175 $ K, in good agreement with the Curie temperature
estimated from neutron data (vide infra). From the Curie constant
we estimate magnetic moment per ion which is $6.01 \mu_B$. This
value is in disagreement with the one theoretically expected for
an appropriate, random mixture of Mn$^{+3}$ and Mn$^{+4}$ ions,
$\mu^2=(1-x)g^2S_3(S_3+1)+xg^2S_4(S_4+1)$, where $S_{3,4}=2$ and
$3/2$ for Mn$^{+3}$ and Mn$^{+4}$ respectively. Enhanced
paramagnetic effective moment has also been observed in Sr-based
compound and has been attributed to magnetic
clusters.\cite{nogues01,nojiri99}
The larger effective magnetic moment of the M$^{/}$ phase for
$x>0.15$ may be related with the idea of magnetic polaron that
"survives" in the regime $T_{\rm C}\le T\le T_{\rm JT}$. We must
note that this state is more relevant when the system moves away
from the CAF regime. Finally, it is interesting to note the
similarity of our bulk magnetic moment data with those of Refs.
\onlinecite{nogues01,nojiri99} referring to the Sr-compound. There
is however a distinct difference below $T_{\rm C}$. Our ac
susceptibility, reduces significantly  for $T< T_{\rm M^/M^{//}}$,
while in the Sr-case it increases in a stepwise fashion. This
"fine" difference may imply that the M$^{//}$ states of Ca and Sr
compounds are different. The orthorhombic strain and the long
in-plane Mn-O2 distance reduction is very similar but the ordered
magnetic moment, bulk magnetic moment and ac susceptibility show
opposite change at $T_{\rm M^/M^{//}}$.

\begin{figure}[tbp] \centering%
\includegraphics[angle=0,width=1.0 \columnwidth]{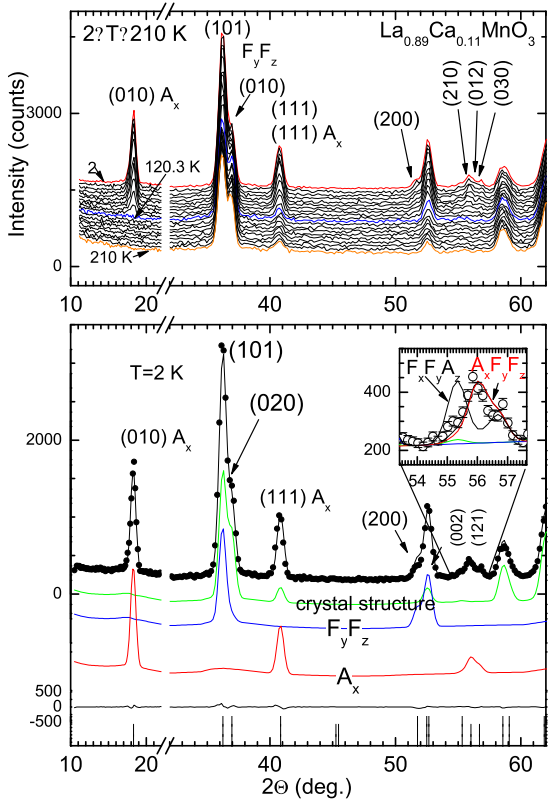}
\caption{ (Upper panel) Selected portion of the neutron powder
diffraction patterns for La$_{0.89}$Ca$_{0.11}$MnO$_{3}$ for $2\le
T \le 210$ K. (Lower panel) Decomposition of the neutron
diffraction pattern as $T=2$ K. Shown are the profiles coming from
the A- antiferromagnetic, the ferromagnetic and crystal
structures. The inset shows the observed and calculated profiles
for the Bragg peaks $(210), (012)$ and $(030)$ based on the models
$F_xF_yA_z$ and $A_xF_yF_z$ . }
\label{m1}%
\end{figure}%

Let us present now the neutron diffraction data related with the
magnetic structure. The magnetic structures for all the samples
are best monitored by observing the neutron diffraction patterns,
using data from E6 instrument with $\lambda=2.4$\AA. Fig. \ref{m1}
show portions of neutron diffraction patterns in the region where
the magnetic coherent intensity is maximum for $x=0.11$ sample.
Similar data were also collected  for the other samples. In all
samples we observed an increasing of the Bragg intensity for the
peaks $(101)$ and $(020)$ for $T<T_{\rm C}$, something which is
compatible with the development of long range ferromagnetic order.
In addition, the samples $x=0.11, 0.125$ and 0.13 show the
characteristic magnetic peaks of the $A$ antiferromagnetic
structure. The most pronounced magnetic Bragg peak of the
A-magnetic structure is the $(010)$ located at $\approx 18^{\rm
o}$. As $x$ increases, the area of this peak at $T=2$ K reduces,
implying that the antiferromagnetic ordered moment reduced and
finally becomes zero for $x\gtrapprox 0.14$.
We refined the $x=0.11$ sample, where the pronounced
orthorhombicity enables us to safely determine the spin components
along the three axes. The lower panel of Fig. \ref{m1} shows a
decomposition of the neutron diffraction pattern at $T=2$ K, which
consists of the patterns coming from the A-type antiferromagnetic,
the ferromagnetic and the crystal structure patterns. The inset of
the lower panel of Fig. \ref{m1} shows the calculated patterns of
models $F_xF_yA_z$ and $A_xF_yF_z$ ($F_x$ denotes ferromagnetic
moment along $x$-axis and $A_x$ magnetic moment along $x$-axis
which follows the A-antiferromagnetic structure). The agreement
between the theoretical  and experimental profile is better for
the $A_xF_yF_z$ model. This magnetic model is closely connected
with the canted magnetic structure observed in the LaMnO$_3$
compound.\cite{moussa96} Similarly, for $x=0.125$ sample, except
for the major ferromagnetic component, small intensity coming from
$A$-type antiferromagnetic structure is present. In addition, the
negligible intensity of the $(200)$ magnetic peak, implies that
the ferromagnetic component cannot be parallel with $b$-axis.
Therefore there are two possibilities. The first is the
antiferromagnetic component to be parallel with the $c$-axis,
while the ferromagnetic component parallel with $a$ and $b$ axis.
The second possibility is both the antiferromagnetic and part of
the ferromagnetic component to be parallel with $a$-axis. Since it
is not possible to have simultaneous ferromagnetic and
antiferromagnetic spin components along the $a$-axis in the same
domain these two different magnetic states concern different
domains. This finding is not strange since in this concentration
regime the phase boundary separating the CAF and FMI ground states
exists.

For the sample $x=0.15$ the NDP at $T=2$ K shows only
ferromagnetic magnetic peaks. The orthorhombicity of the sample
permits the calculation of the orientation of the magnetic moment
in respect of the crystallographic axes. Thus, the case ${\bf
m}||b$-axis should be excluded. If the magnetic moment was along
the $b$-axis the magnetic contribution at the $(020)$ peak should
be zero. Based on the best agreement between observed and
theoretically calculated profiles we conclude that the ordered
magnetic moment should be lie on the $ac$-plane with its larger
component along the $a$-axis. Nevertheless, a small component
along the $b$-axis cannot be totally excluded. The ordered
magnetic moments as estimated from Rietveld refinements for the
$x=0.13, 0.15$ and 0.175 samples are plotted in the lower panels
of  Figs. \ref{abc-r013}, \ref{abc-r015} and \ref{abc-r0175},
respectively.
A weak but distinctive anomaly is present in the temperature
variation curves of the ordered magnetic moments at $T_{\rm
M^/M^{//}}$. It is imperative to note that for $T\approx T_{\rm
M^/M^{//}}$ the ordered magnetic moment curve, for $x=0.15$ and
0.175 samples, displays a small slope change.  For samples
$x=0.11, 0.125$ and 0.13 where we can detect the magnetic peak
from the $A-$type magnetic structure, this anomalous behavior
certainly is related with the antiferromagnetic component.
For $x\ge 0.15$  it is possible that a small percentage of the
spin of Mn ions may prefer a glass like behavior for $T<T_{\rm
M^/M^{//}}$, as the ac susceptibility measurements
show.\cite{markovich02a} Concerning this "strange" magnetic
component it is not clear how it is connected with the structural
transition at the same temperature. One could invoke the idea of
the phase separation to explain this behavior, however, the bulk
character of the concomitant structural transition at $T_{\rm
M^/M^{//}}$ is not so favorable for this interpretation. Based on
the experimental data one may say that the structural transition
prevents some part of the Mn-spin to participate in the long-range
ordering.\cite{markovich02a,pissas04a} This structural transition
should be related with the orbital degree of freedom in agreement
with M\"{o}ssbauer\cite{pissas04b} and NMR\cite{papavassiliou03}
data.

\section{Discussion}

Unexpectedly, the space group $P 2_1/c$ has been used by Lobanov
et al.,\cite{lobanov00} in order to describe the crystal structure
for a sample with nominal stoichiometry
La$_{0.85}$Ca$_{0.15}$MnO$_3$ prepared by liquid road and final
reaction temperature 900$^{\rm o}$C. However, although this space
group is the same with that used in our study, several differences
exist. More specifically: (a) The ac susceptibility and
resistivity data are different. (b) The sample of Lobanov et
al.,\cite{lobanov00} does not show the pseudo-orthorhombic
splitting seen in our data and in the La$_{1-x}$Sr$_x$MnO$_3$
compound. The lattice parameters follow a monotonic temperature
variation, contrary to ours. (c) The particular arrangement of the
short bonds of Mn1-O and Mn2-O is different with it deduced from
our data. Since no more data, from samples prepared with method of
Ref. \onlinecite{lobanov00}, are available, we cannot safely
further comment on the possible relation between the two studies.

It is interesting to discuss our crystal structure data in terms
of the normal coordinates\cite{kanamori60} of the octahedral
complex MnO$_6$,  $Q_3$ and $Q_2$. An arbitrary $e_g$ orbital
state can be written as
$|\theta\rangle=\cos(\theta/2)|d_{z^2}\rangle+\sin(\theta/2)|d_{x^2-y^2}\rangle$.
The orbitals $|d_{z^2}\rangle$ and $|d_{x^2-y^2}\rangle$
correspond to the states $|\theta=0\rangle$ and
$|\theta=\pi\rangle$ respectively. It is also possible, the
orbitals ($|d_{x^2}\rangle$, $|d_{y^2}\rangle$) and (
$|d_{z^2-x^2}\rangle$, $|d_{z^2-y^2}\rangle$) to be formed. In
this case we have $|\theta=\pm (2/3)\pi\rangle$ and $|\theta=\pm
(1/3)\pi\rangle$, respectively.   Experimentally the orbital
filling can be estimated from crystallographic data, through the
Mn-O bond lengths using the equation\cite{kanamori60,kugel82}
$\tan\theta=Q2/Q3$, where $Q_3=2/\sqrt{6}(2m-l-s)$  and
$Q_2=\sqrt{2}(l-s)$. Here $l,m$ and $s$ mean long, medium and
short Mn-O bond lengths, respectively. $Q_2$
 mode mixes the $|d_{x^2-y^2}\rangle$
and $|d_{z^2}\rangle$ orbitals and produces a two sublattice
orbital ordering pattern, with a lower energy orbital
$\Psi_g=c_1|d_{x^2-y^2}\rangle\pm c_2|d_{z^2}\rangle$ and a higher
energy orbital $\Psi_e=c_2|d_{x^2-y^2}\rangle\mp
c_1|d_{z^2}\rangle$ ($\pm$ is referring to the two sublattices).

\begin{figure*}[tbp] \centering%
\includegraphics[angle=0,width=5 in]{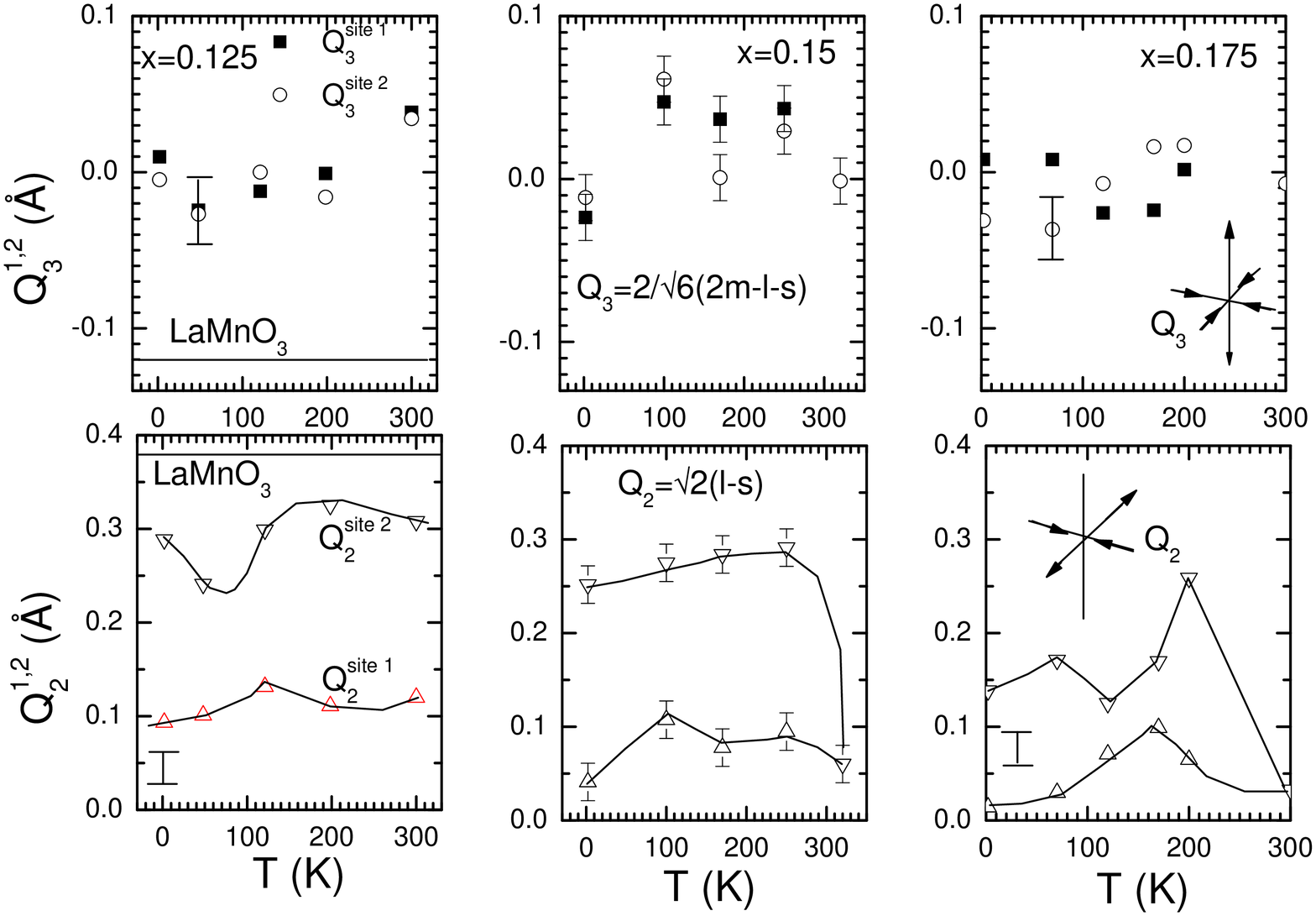}
\caption{ Temperature variation of the $Q_2$ and $Q_3$ parameters
for La$_{1-x}$Ca$_x$ MnO$_3$ ($x=0.125, 0.15$ and 0.175) samples.
The data for the LaMnO$_3$ compound are also included for a direct
comparison (thin line).}
\label{q2q3}%
\end{figure*}

Using the crystallographic data at $T=300$ K of LaMnO$_3$ (Ref.
\onlinecite{carvajal98}) one finds, $Q_3=-0.121$\AA\  and
$Q_2=0.383$\AA. These values correspond to a canted
antiferromagnet like orbital arrangement with two sublattices,
with $\theta=\pm 72.38^{\rm o}$, respectively. Based on the
crystallographic data of the present study we calculated the $Q_2$
and $Q_3$ parameters for the two cites predicted from $P 2_1/c$
structural model. Fig. \ref{q2q3} shows the temperature variation
of $Q_2$ and $Q_3$ for the $x=0.125, 0.15$ and 0.175 samples. With
solid lines we also plot the values which correspond to the
LaMnO$_3$ compound. The first comment which can be made is that
both $Q_2$ and $|Q_3|$ parameters for the two sites have lower
values, in comparison to these of the LaMnO$_3$ compound. The
second one is that, both Mn sites have $Q_3\approx 0 $ with the
first displaying a high value while the second a moderate one.
Since we only have  a small number of data it is difficult to
recognize any change at $T_{\rm M^/M^{//}}$. The third comment is
that, only the sample $x=0.125$ shows non monotonous variation of
$Q_2$ at $T_{\rm M^/M^{//}}$. We note that for $T>T_{\rm JT}$ the
two sites pattern collapse into one site, with both $Q_2$ and
$Q_3$ modes takng values near zero.

\subsection{The comparison with Sr-compound}

First of all, we must point out the similarities between the Sr
and Ca based compounds, by comparing the neutron data from the
stoichiometric La$_{1-x}$Sr$_x$MnO$_3$ ($0.1\le x\le 0.2$) of
Dabrowski et al.\cite{dabrowski99,xiong99}with ours. As Dabrowski
et al. have pointed out,\cite{dabrowski99} the phase diagrams of
the Sr case reported in the recent literature were constructed
using physical and structural information obtained for dissimilar
samples. Specifically, in the doping regime $0\le x<0.3$ of
Sr-compound the structure can form vacancies on both the A(La,Sr)
and B(Mn) sites during synthesis under oxidizing conditions.
Formation of equal number of vacancies on the A and B sites would
increase the average Mn oxidation state. Let us discuss our
results with those of single crystal data of
La$_{7/8}$Sr$_{1/8}$MnO$_3$ compound (this concentration is the
most representative and widely studied of the particular regime)
where very interesting results are available.

In a recent resonant x-ray scattering study on a
La$_{7/8}$Sr$_{1/8}$MnO$_{3}$ single crystal, Geck et
al.,\cite{geck04} have shown that the orbital state occurring in
the interval $T_{\rm MT}\le T\le T_{JT}$, (M means monoclinic and
T triclinic) is similar to that found in LaMnO$_3$. For $T<T_{\rm
MT}$ an orbital reordering is compatible with the resonant data.
This group\cite{geck04} observed non zero resonant and nonresonant
intensity at (3/2, 3/2, 3) superlattice reflection, only for
$T<T_{\rm MT}$. The nonresonant scattering in the particular
reflection implies some kind of a superstructure modulation that
develops at $T=T_{\rm MT}$. Such behavior is related to structural
and electronic degrees of freedom, speculatively attributed to the
orbital polarons formation.\cite{mizokawa00,kilian98}
The work of Geck et al.,\cite{geck04} further supports the aspect
that the CAF and FMI phases of Ca and Sr based
La$_{1-x}$(Ca,Sr)$_x$MnO$_3$ compounds have both common
characteristics, and distinct differences. It seems that the
vertical boundary of the Sr-case that separates the O$^/$ CAF
ground state from the triclinic new unknown orbital state exists
even in the Ca-case. Our results show that: (a) Similar to the Sr
compound for $T_{\rm MT}\le T\le T_{\rm JT}$ the structure is
monoclinic $(P 2_1/c)$. (b) Based on our high resolution
synchrotron x-ray diffraction data, the crystal structure of the
Ca case remains monoclinic, while in the Sr case it is triclinic
and (c) both families of compounds exhibit significantly lower JT
distortion in comparison with the M$^/$ phase which occurs for
$T_{\rm M^/M^{//}}<T<T_{\rm TJ}$.

\subsection{Comparison between stoichiometric and cation deficient samples}

 We would like to emphasize once again
that our results concern stoichiometric samples. The
stoichiometric samples have properties different from those
prepared in air. \cite{huang98,pissas04a,dai96} The most prominent
differences between the two families of samples are that the
oxygen deficient samples do not display the CAF structure and do
not undergo the characteristic cooperative JT distortion,
irrespectively of the nominal calcium concentration. The second
difference makes the cation deficient samples to be in an orbital
liquid state below $T_{\rm C}$, in the FMI regime, contrary to the
stoichiometric ones where the ferromagnetic phase takes place in
the cooperative ordered JT state at least in the interval $T_{\rm
M^/M^{//}}<T<T_{\rm C}$. In addition, the cation deficient samples
(e.g. $x=0.1, 0.15$ samples\cite{huang98,dai96b}) show an almost
monotonous variation with temperature of the unit cell parameters,
satisfying also the inequality $c>a>b/\sqrt{2}$ (for $T<T_{\rm
C}$), contrary to the behavior observed in our stoichiometric
samples where a pronounced non monotonous variation is observed.
It is now well established that for low Ca concentrations the air
prepared samples have cation
vacancies.\cite{pissas04a,dabrowski99a,roosmalen94} The amount of
vacancies increases as the nominal Ca concentration is near $x=0$.
These samples are ferromagnetic irrespectively of $x$. The ordered
magnetic moment at zero temperature is significantly lower than
the corresponding moment of the stoichiometric samples. For the
$x=0.1$ and 0.15 non stoichiometric samples the ordered magnetic
moment was found $\sim 3 \mu_{\rm B}$ (Refs.
\onlinecite{huang98,dai96b,algarabel03}), while in our
stoichiometric samples are obtained $\sim 3.7\mu_{\rm B}$.

\begin{figure}[tbp] \centering%
\includegraphics[angle=0,width=0.7 \columnwidth]{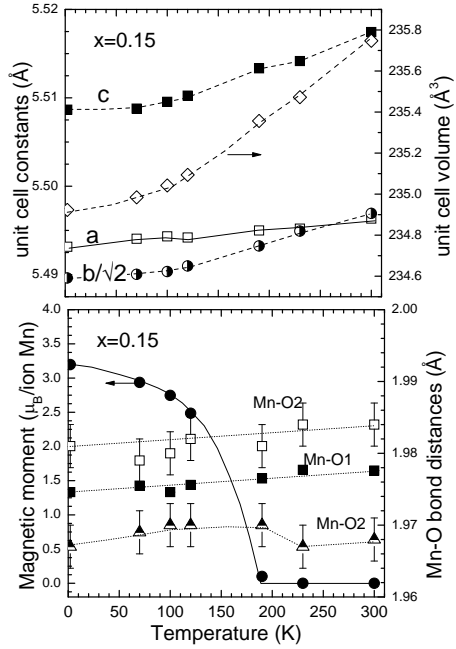}
\caption{ (upper panel) Temperature variation of the lattice
parameters and unit-cell volume for
La$_{0.85}$Ca$_{0.15}$MnO$_{3}$ compound prepared in air
atmosphere. (lower panel) Temperature variation of the ordered
ferromagnetic moment and Mn-O$_i$ bond lengths.  Lines through the
experimental points are guides to the eye.}
\label{ap015}%
\end{figure}

In order to check once more the situation we selected the $x=0.15$
sample, prepared in air for a detailed structural characterization
using neutron diffraction data. Fig. \ref{ap015} shows the
temperature variation of the unit cell parameters (upper panel),
the ordered magnetic moment and the Mn-O bond lengths (lower
panel), as they were deduced from neutron diffraction data using
E9 instrument. In perfect agreement with the above, this sample is
ferromagnetic insulator without the orthorhombic splitting that
characterizes the cooperative JT distortion of the reduced
samples. Interestingly, the ordered magnetic moment is lower than
the one of the reduced samples (e.g 3.2 vs 3.7$\mu_B$).

In the work of Algarabel et al.\cite{algarabel03} that concerns a
cation deficient sample with nominal $x=0.1$, significant small
angle neutron scattering intensity ( due to nanometric clusters)
has been found to coexist with the long-range ferromagnetic phase.
In our opinion, it is difficult to correlate this nanometric
clusters with the anomalies observed for $T<T_{\rm M^/M^{//}}$ in
the stoichiometric samples because these clusters may be related
with cation vacancies. It seems that in samples prepared in air
the ratio $R(P_{\rm O_2},T)= \rm{Mn}^{+4}/\rm{Mn}^{+3}$ is
controlled by the partial oxygen pressure. When these samples are
cooled slowly during the final preparation step the $R(P_{\rm
O_2},T)$ is different from the nominal Ca concentration,
$R_x=x/(1-x)$.
In order the electric charge neutrality to be maintained, cation
vacancies are created.
This effect is significant for $x<0.2$. Therefore, in order to
produce stoichiometric samples we must use reduced atmosphere
conditions or  quench the sample from high temperature to liquid
N$_2$.\cite{pissas04a,dabrowski99a,roosmalen94} The presence of
cation vacancies completely destroys  the coherent JT distortion
even for the $x=0$ sample.\cite{pissas04a} In the cation deficient
samples firstly, it seems that the $R= \rm{Mn}^{+4}/\rm{Mn}^{+3}$
is locked to a nearly  independent $x$ value, and secondly the
M$^/$ phase which concerns a coherent JT distortion cannot be
formed due to the cation vacancies. In this context  numerous
experiment results in the particular Ca regime should be
interpreted. It is a mystery why the cation deficient samples show
similar bulk magnetic measurements near $T=100$ K as the
stoichiometric ones. The stoichiometric samples display a
structural transition at $T_{\rm M^/M^{//}}$, therefore it is
reasonable to correlate this anomaly with the particular shape of
the curves obtained in dc magnetic
 and ac-susceptibility measurements. For non stoichiometric samples such
structural anomaly is absent leaving open the question about the
relation of the bulk magnetization anomaly among the two families
of samples.

Summarizing, the crystal structure refinements , from
stoichiometric samples of the La$_{1-x}$Ca$_x$MnO$_3$ compound in
the FMI regime, are presented. All the samples ($0.11\le x\le
0.175$) undergo a structural transition from O* to M$^/$ structure
at $T_{\rm JT}$. High resolution x-ray diffraction data revealed
that for $T<T_{\rm JT}$, La$_{1-x}$Ca$_x$MnO$_3$ compounds
($0.11\le x\le 0.175$) have monoclinic $P 2_1/c$ symmetry.
The major structural characteristic of this model is the two Mn
sites. As the bond lengths are concerned the Mn2 site displays
properties similar with those of the Mn site of the LaMnO$_3$
compound, while the Mn1 site, resembles the Mn site in the FMM
regime. By further cooling, below the characteristic temperature
$T_{\rm M^/M^{//}}$, although, the structure remains the same, a
strong reduction of the strain parameter $s=(a-c)/(a+c)$ is
observed. This change has been accompanied by significant
reduction of the monoclinic $\gamma$ angle and is correlated with
distinct changes in dc and ac magnetic measurements.

\acknowledgements{This work was partial supported from EU through
the CHRX-CT93-0116 access to large-scale facilities project. }


\begin{references}
\bibitem{monoclinic} As the present study reveals that the true
structure below the $T_{\rm JT}$ is monoclinic, we prefer to keep
the symbols O$^/$ and O$^{//}$ in order to help the reader to
compare our results with those of other studies.
\end{references}
\end{document}